\documentclass[journal,12pt,twocolumn]{IEEEtran}
\usepackage{graphicx}
\usepackage{diagbox}
\usepackage{color} 
\usepackage{caption}
\usepackage{subcaption}
\usepackage{etoolbox}
\usepackage{amssymb,amsmath,amsfonts,amsthm,stackengine}
\usepackage{tikz}
\usepackage{epstopdf}
\usepackage[affil-it]{authblk}
\usepackage{lipsum}
\theoremstyle{plain}
\newtheorem{assumption}{Assumption}

\newtoggle{doublecolumn}
\newtoggle{blindreview}
\togglefalse{blindreview}
\begin{document}

\title{WiSpeed: A Statistical Electromagnetic Approach for Device-Free Indoor Speed Estimation}


\author[${\star \dagger}$]{Feng Zhang}
\author[${\star \dagger}$]{Chen Chen}
\author[${\star \dagger}$]{Beibei Wang}
\author[${\star \dagger}$]{K. J. Ray Liu}

\affil[${\star}$]{University of Maryland, College Park, MD 20742, USA.}
\affil[${\dagger}$]{Origin Wireless, Inc., 7500 Greenway Center Drive, MD 20770, USA.}
\affil[$\star$]{Email: \{fzhang15, cc8834, bebewang, kjrliu\}@umd.edu}

\maketitle

\begin{abstract}
Due to the severe multipath effect, no satisfactory device-free methods have ever been found for indoor speed estimation problem, especially in non-line-of-sight scenarios, where the direct path between the source and observer is blocked. In this paper, we present WiSpeed, a universal low-complexity indoor speed estimation system leveraging radio signals, such as commercial WiFi, LTE, 5G, etc., which can work in both device-free and device-based situations. By exploiting the statistical theory of electromagnetic waves, we establish a link between the autocorrelation function of the physical layer channel state information and the speed of a moving object, which lays the foundation of WiSpeed. WiSpeed differs from the other schemes requiring strong line-of-sight conditions between the source and observer in that it embraces the rich-scattering environment typical for indoors to facilitate highly accurate speed estimation. Moreover, as a calibration-free system, WiSpeed saves the users' efforts from large-scale training and fine-tuning of system parameters. In addition, WiSpeed could extract the stride length as well as detect abnormal activities such as falling down, a major threat to seniors that leads to a large number of fatalities every year. Extensive experiments show that WiSpeed achieves a mean absolute percentage error of $4.85\%$ for device-free human walking speed estimation and $4.62\%$ for device-based speed estimation, and a detection rate of $95\%$ without false alarms for fall detection.
\end{abstract}

\section{Introduction}
\label{sec:Introduction}
As people are spending more and more their time indoors nowadays, understanding their daily indoor activities will become a necessity for future life. Since the speed of the human body is one of the key physical parameters that can characterize the types of human activities, speed estimation of human motions is a critical module in human activity monitoring systems. Compared with traditional wearable sensor-based approaches, device-free speed estimation is more promising due to its better user experience, which can be applied in a wide variety of applications, such as smart homes~\cite{khan16IoT}, health care~\cite{Pinto17Wecare}, fitness tracking~\cite{Schaefer16Fitness}, and entertainment.

Nevertheless, indoor device-free speed estimation is very challenging mainly due to the severe multipath propagations of signals and the blockage between the monitoring devices and the objects under monitoring. Conventional approaches of motion sensing require specialized devices, ranging from RADAR, SONAR, laser, to camera. Among them, the vision-based schemes~\cite{Wang10Machine} can only perform motion monitoring in their fields of vision with performance degradation in dim light conditions. Also, they introduce privacy issues. Meanwhile, the speed estimation produced by RADAR or SONAR~\cite{Gurbuz16Micro} varies for different moving directions, mainly because of the fact that the speed estimation is derived from the Doppler shift which is relevant to the moving direction of an object. Also, the multipath propagations of indoor spaces further undermine the efficacy of RADAR and SONAR.

More recently, WiGait~\cite{Hsu17Gait} and WiDar~\cite{Qian17WiDar} are proposed to measure gait velocity and stride length in indoor environments using radio signals. However, WiGait uses specialized hardware to send Frequency Modulated Carrier Wave (FMCW) probing signals, and it requires a bandwidth as large as $1.69\,$GHz to resolve the multipath components. On the other hand, WiDar can only work well under a strong line-of-sight (LOS) condition and a dense deployment of WiFi devices since its performance relies heavily on the accuracy of ray tracing/geometry techniques.

In this paper, we present WiSpeed, a robust universal speed estimator for human motions in a rich-scattering indoor environment, which can estimate the speed of a moving object under either the device-free or device-based condition. WiSpeed is actually a fundamental principle which requires no specific hardware as it can simply utilize only a single pair of commercial off-the-shelf WiFi devices. First, we characterize the impact of motions on the autocorrelation function (ACF) of the received electric field of electromagnetic (EM) waves using the statistical theory of EM waves. However, the received electric field is a vector and it cannot be easily measured. Therefore, we further derive the relation between the ACF of the power of the received electric field and the speed of motions, since the electric field power is directly measurable on commercial WiFi devices~\cite{Halperin11tool}. By analyzing different components of the ACF, we find that the first local peak of the ACF differential contains the crucial information of speed of motions, and we propose a novel peak identification algorithm to extract the speed. Furthermore, the number of steps and the stride length can be estimated as a byproduct of the speed estimation. In addition, fall can be detected from the patterns of the speed estimation.

To assess the performance of WiSpeed, we conduct extensive experiments in two scenarios, namely, human walking monitoring and human fall detection. For human walking monitoring, the accuracy of WiSpeed is evaluated by comparing the estimated walking distances with the ground-truths. Experimental results show that WiSpeed achieves a mean absolute percentage error (MAPE) of $4.85\%$ for the case when the human does not carry the device and a MAPE of $4.62\%$ for the case when the subject carries the device. In addition, WiSpeed can extract the stride lengths and estimates the number of steps from the pattern of the speed estimation under the device-free setting. In terms of human fall detection, WiSpeed is able to differentiate falls from other normal activities, such as sitting down, standing up, picking up items, and walking. The average detection rate is $95\%$ with no false alarms. To the best of our knowledge, WiSpeed is the first device-free/device-based wireless speed estimator for motions that achieves high estimation accuracy, high detection rate, low deployment cost, large coverage, low computational complexity, and privacy preserving at the same time.

Since WiFi infrastructure is readily available for most indoor spaces, WiSpeed is a low-cost solution that can be deployed widely. WiSpeed would enable a large number of important indoor applications such as
\begin{enumerate}
\item \textit{Indoor fitness tracking:} More and more people become aware of their physical conditions and are thus interested in acknowledging their amount of exercise on a daily basis. WiSpeed can assess a person's exercise amount by the estimation of the number of steps through the patterns of the speed estimation. With the assistance of WiSpeed, people can obtain their exercise amount and evaluate their personal fitness conditions without any wearable sensors attached to their bodies.
\item \textit{Indoor navigation:} Although outdoor real-time tracking has been successfully solved by GPS, indoor tracking still leaves an open problem up to now. Dead reckoning based approach is among the existing popular techniques for indoor navigation, which is based upon measurements of speed and direction of movement to compute the position starting from a reference point. However, the accuracy is mainly limited by the inertial measurement unit (IMU) based moving distance estimation. Since WiSpeed can also measure the speed of a moving WiFi device, the accuracy of distance estimation module in dead reckoning-based systems can be improved dramatically by incorporating WiSpeed.
\item \textit{Fall detection:} Real-time speed monitoring for human motions is important to the seniors who live alone in their homes, as the system can detect falls which impose major threats to their lives.
\item \textit{Home surveillance:} WiSpeed can play a vital role in the home security system since WiSpeed can distinguish between an intruder and the owner's pet through their different patterns of moving speed and inform the owner as well as the law enforcement immediately.
\end{enumerate}

The rest of the paper is organized as follows. Section \ref{sec:Related works} summarizes the related works about human activity recognition using WiFi signals. Section \ref{sec:Statistical theory} introduces the statistical theory of EM waves in cavities and its extensions for wireless motion sensing. Section \ref{sec: Theoretical Foundation of WiSpeed} presents the basic principles of WiSpeed and Section \ref{sec: Key components of WiSpeed} shows the detailed designs of WiSpeed. Experimental evaluation is shown in Section \ref{sec: Exp evaluations}. Section \ref{sec:discussion} discusses the parameter selections and the computational complexity of WiSpeed and Section \ref{sec: Conclusions} concludes the paper.

\section{Related Works}
\label{sec:Related works}
Existing works on device-free motion sensing techniques using commercial WiFi include gesture recognition~\cite{Abdelnasser15WiGest,Qian17Inferring,Ali15Keystrode,Pu13Whole,Wang16WiHear}, human activity recognition~\cite{Wang14Eeyes,Wang15Understanding,Han14WiFall}, motion tracing~\cite{Sun15WiDraw,Adib13See}, passive localization~\cite{Qian17WiDar,Seifeldin13Nuzzer}, vital signal estimation~\cite{Chen17TRBreath}, indoor event detection~\cite{Xu17TRIEDS} and so on. These approaches are built upon the phenomenon that human motions inevitably distort the WiFi signal and can be recorded by WiFi receivers for further analysis. In terms of the principles, these works can be divided into two categories: learning based and ray-tracing based. Details of the two categories are elaborated below.

\textit{Learning-based:}
These schemes consist of two phases, namely, an offline phase, and an online phase. During the offline phase, features associated with different human activities are extracted from the WiFi signals and stored in a database; in the online phase, the same set of features are extracted from the instantaneous WiFi signals and compared with the stored features so as to classify the human activities. The features can be obtained either from CSI or the Received Signal Strength Indicator (RSSI), a readily available but low granularity information encapsulating the received power of WiFi signals. For example,  E-eyes~\cite{Wang14Eeyes} utilizes histograms of the amplitudes of CSI to recognize daily activities such as washing dishes and brushing teeth. CARM~\cite{Wang15Understanding} exploits features from the spectral components of CSI dynamics to differentiate human activities. WiGest~\cite{Abdelnasser15WiGest} exploits the features of RSSI variations for gesture recognition.

A major drawback of the learning-based approach lies in that these works utilize the speed of motion to identify different activities, but they only obtain features related to speed instead of directly measuring the speed. One example is the Doppler shift, as it is determined by not only the speed of motion but also the reflection angle from the object as well. These features are thus susceptible to the external factors, such as the changes in the environment, the heterogeneity in human subjects, the changes of device locations, etc., which might violate their underlying assumption of the reproducibility of the features in the offline and online phases.

\textit{Ray-tracing based:}
Based on the adopted techniques, they can be classified into multipath-avoidance and multipath-attenuation. The multipath-avoidance schemes track the multipath components only reflected by a human body and avoid the other multipath components. Either a high temporal resolution~\cite{Adib14WiTrack} or a ``virtual'' phased antenna array is used~\cite{Adib13See}, such that the multipath components relevant to motions can be discerned in the time domain or in the spatial domain from those irrelevant to motions. The drawback of these approaches is the requirement of dedicated hardware, such as USRP, WARP~\cite{murphy2006design}, etc., to achieve a fine-grained temporal and spatial resolution, which is unavailable on WiFi devices~\footnote{On commercial main-stream 802.11ac WiFi devices, the maximum bandwidth is $160\,$MHz, much smaller than the $1.69\,$GHz bandwidth in WiTrack. Meanwhile, commercial WiFi devices with multiple antennas cannot work as a (virtual) phased antenna array out-of-box before carefully tuning the phase differences among the RF front-ends.}.

In the multipath-attenuation schemes, the impact of multipath components is attenuated by placing the WiFi devices in the close vicinity of the monitored subjects, so that the majority of the multipath components are affected by the subject~\cite{Qian17WiDar,Qian17Inferring,Sun15WiDraw}. The drawback is the requirement of a very strong LOS working condition, which limits their deployment in practice.

WiSpeed differs from the state-of-the-arts in literature in the following ways:
\begin{itemize}
\item WiSpeed embraces multipath propagations indoors and can survive and thrive under severe non-line-of-sight (NLOS) conditions, instead of getting rid of the multipath effect~\cite{Qian17WiDar,Qian17Inferring,Adib13See,Adib14WiTrack}.
\item WiSpeed exploits the physical features of EM waves associated with the speed of motion and estimates the speed of motion without detouring. As the physical features hold for different indoor environments and human subjects, WiSpeed can perform well disregarding the changes of environment and subjects and it is free from any kind of training or calibration.
\item WiSpeed enjoys its advantage in a lower computational complexity in comparison with other approaches since costly operations such as principal component analysis (PCA), discrete wavelet transform (DWT), and short-time Fourier transform (STFT)~\cite{Qian17WiDar,Ali15Keystrode,Wang15Understanding} are not required.
\item WiSpeed is a low-cost solution since it only deploys a single pair of commercial WiFi devices, while \cite{Hsu17Gait,Qian17WiDar,Pu13Whole,Sun15WiDraw,Adib14WiTrack} need either specialized hardware or multiple pairs of WiFi devices.
\end{itemize}

\section{Statistical Theory of EM Waves for Wireless Motion Sensing}
In this section, we first decompose the received electric field at the Rx into different components and then, the statistical behavior of each component is analyzed under certain statistical assumptions.
\label{sec:Statistical theory}
\subsection{Decomposition of the Received Electric Field}
\begin{figure}[tb]
    \centering
    \begin{subfigure}[b]{0.23\textwidth}
        \includegraphics[width=\textwidth]{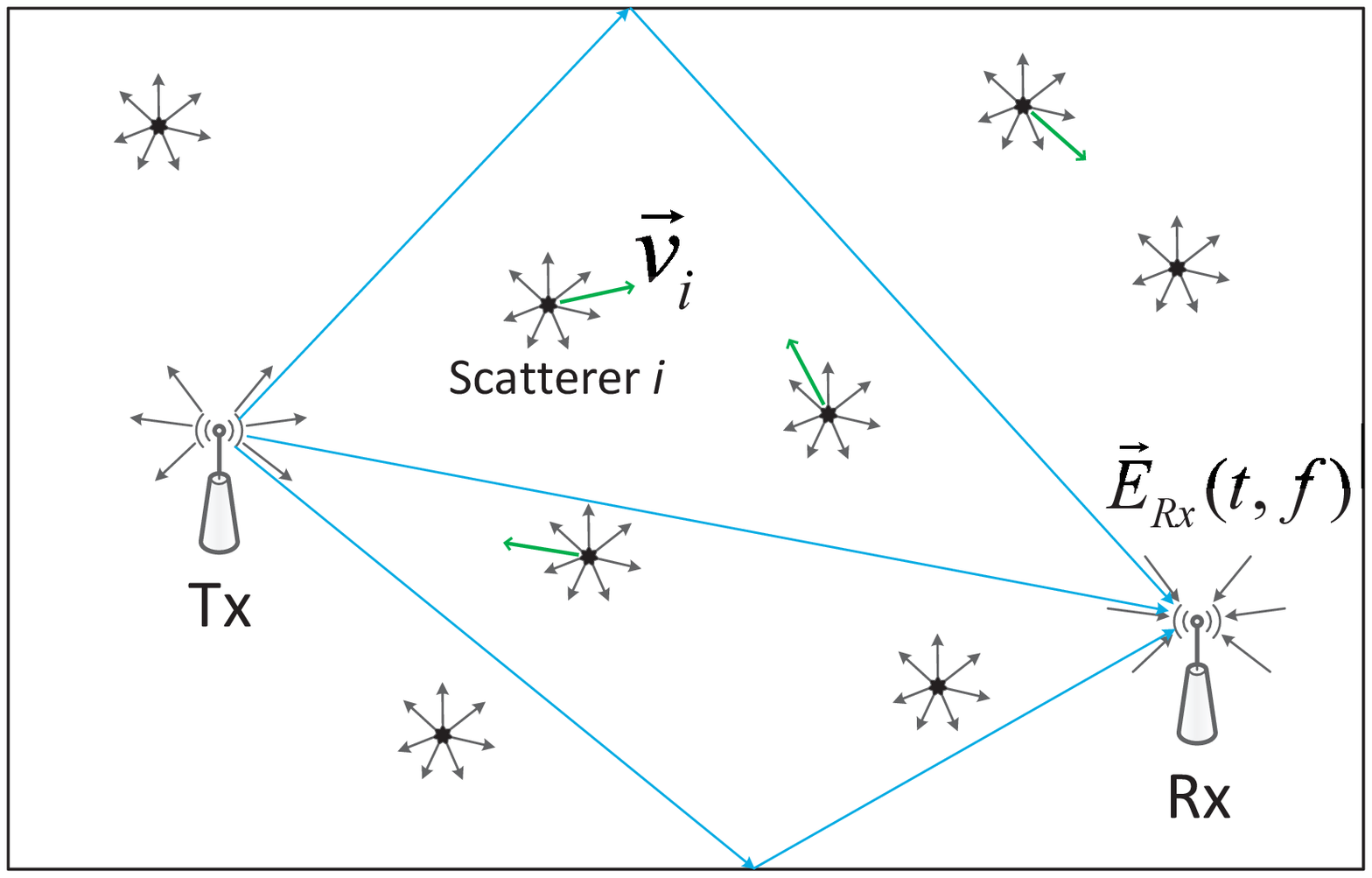}
        \caption{\label{fig:PhysicalExplanations1} Propagation of radio signals in rich scattering environment.}
    \end{subfigure}
    ~
    \begin{subfigure}[b]{0.23\textwidth}
        \includegraphics[width=\textwidth]{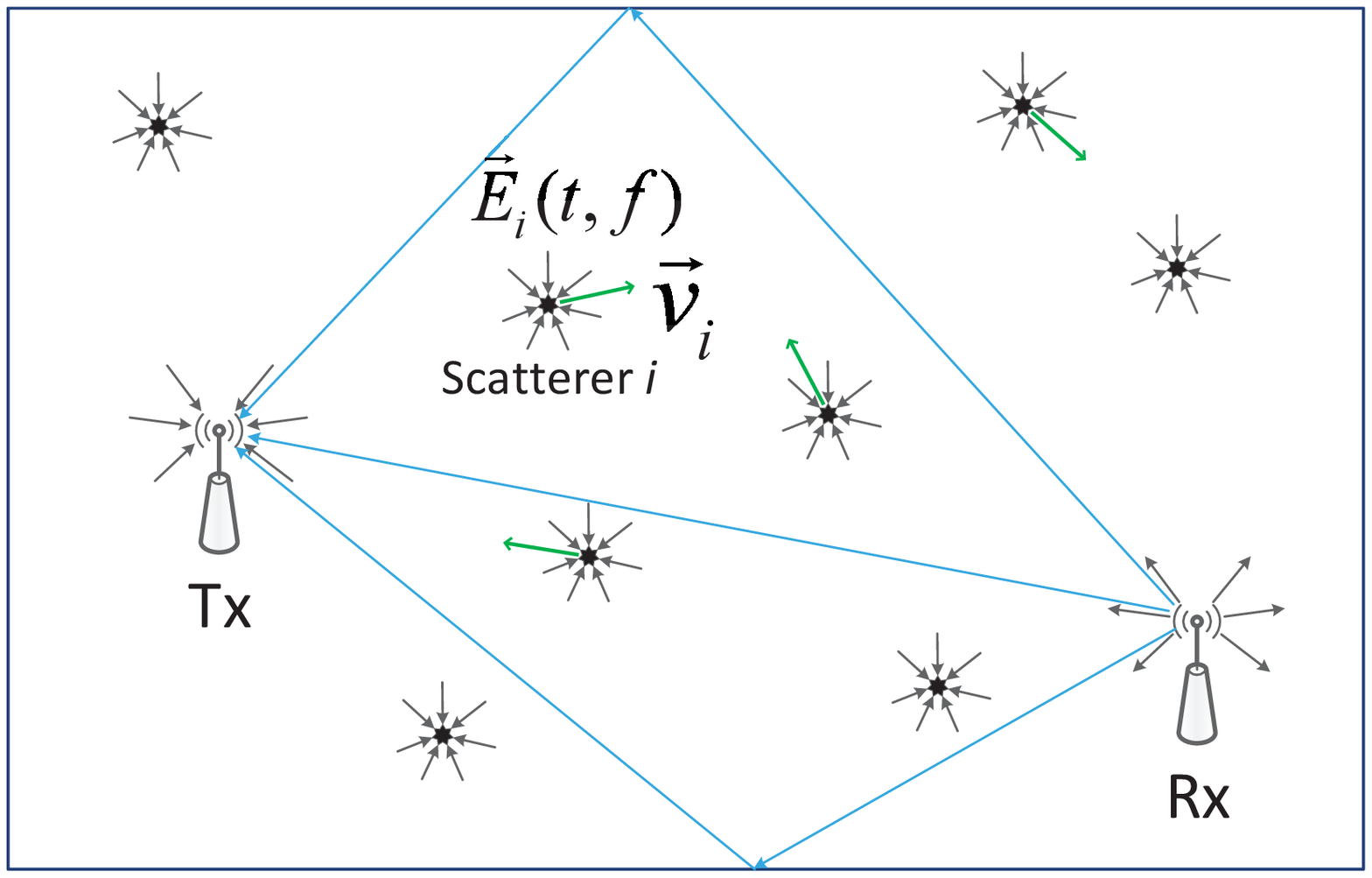}
        \caption{\label{fig:PhysicalExplanations2} Understanding $\vec{E}_i(t,f)$, $i\in\Omega_d(t)$ using channel reciprocity.}
    \end{subfigure}
    \caption{Illustration of wave propagation with many scatterers.}
\end{figure}

To provide an insight into the impact of motions on the EM waves, we consider a rich-scattering environment as illustrated in Fig.~\ref{fig:PhysicalExplanations1}, which is typical for indoor spaces. The scatterers are assumed to be diffusive and can reflect the impinging EM waves towards all directions. A transmitter (Tx) and a receiver (Rx) are deployed in the environment, both equipped with omnidirectional antennas. The Tx emits a continuous EM wave via its antennas, which is received by the Rx. In an indoor environment or a reverberating chamber, the EM waves are usually approximated as plane waves, which can be fully characterized by their electric fields. Let $\vec{E}_{Rx}(t,f)$ denote the electric field received by the receiver at time $t$, where $f$ is the frequency of the transmitted EM wave. In order to analyze the behavior of the received electric field, we decompose $\vec{E}_{Rx}(t,f)$ into a sum of electric fields contributed by different scatterers based on the superposition principle of electric fields
\begin{eqnarray}
\label{eq:E_RX decomposition}
\vec{E}_{Rx}(t,f) = \sum_{i\in\Omega_s(t)} \vec{E}_i(t,f) + \sum_{j\in\Omega_d(t)} \vec{E}_j(t,f)£¬
\end{eqnarray}
where $\Omega_s(t)$ and $\Omega_d(t)$ denote the set of static scatterers and dynamic (moving) scatterers, respectively, and $\vec{E}_i(t,f)$ denotes the part of the received electric field scattered by the $i$-th scatterer. The intuition behind the decomposition is that each scatterer can be treated as a ``virtual antenna'' diffusing the received EM waves in all directions and then these EM waves add up together at the receive antenna after bouncing off the walls, ceilings, windows, etc. of the building. When the transmit antenna is static, it can be considered to be a ``special'' static scatterer, i.e., $Tx\in\Omega_s(t)$; when it is moving, it can be classified in the set of dynamic scatterers, i.e., $Tx\in\Omega_d(t)$. The power of $\vec{E}_{Tx}(t,f)$ dominates that of electric fields scattered by scatterers.

Within a sufficiently short period, it is reasonable to assume that both the sets $\Omega_s(t)$, $\Omega_d(t)$ and the electric fields $\vec{E}_i(t,f)$, $i\in\Omega_s(t)$ change slowly in time. Then, we have the following approximation:
\begin{eqnarray}
\label{eq: E_RX approximated decomposition}
\vec{E}_{Rx}(t,f) \approx \vec{E}_s(f) + \sum_{j\in\Omega_d} \vec{E}_j(t,f),
\end{eqnarray}
where $\vec{E}_s(f) \approx \sum_{i\in\Omega_s(t)} \vec{E}_i(t,f)$.

\subsection{Statistical Behaviors of the Received Electric Field}

\begin{figure}[tb]
	\centering
   \includegraphics[width=0.35\textwidth]{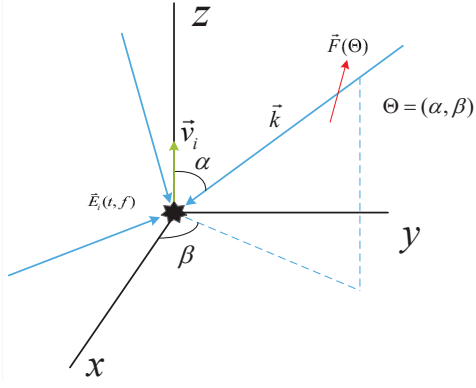}
    \caption{\label{fig:Sphere_Scattering} Plane wave component $\vec{F}(\Theta)$ of the electric field with vector wavenumber $\vec{k}$.}
\end{figure}

As is known from the channel reciprocity, EM waves traveling in both directions will undergo the same physical perturbations (i.e. reflection, refraction, diffraction, etc.). Therefore, if the receiver were transmitting EM waves, all the scatterers would receive the same electric fields as they contribute to $\vec{E}_{Rx}(t,f)$, as shown in Fig.~\ref{fig:PhysicalExplanations2}. Therefore, in order to understand the properties of $\vec{E}_{Rx}(t,f)$, we only need to analyze its individual components $\vec{E}_i(t,f)$, which is equal to the received electric field by the $i$-th scatterer as if the Rx were transmitting. Then, $\vec{E}_{i}(t,f)$ can be interpreted as an integral of plane waves over all direction angles, as shown in Fig.~\ref{fig:Sphere_Scattering}. For each incoming plane wave with direction angle $\Theta=(\alpha,\beta)$, where $\alpha$ and $\beta$ denote the elevation and azimuth angles, respectively, let $\vec{k}$ denote its vector wavenumber and let $\vec{F}(\Theta)$ stand for its angular spectrum which characterizes the electric field of the wave. The vector wavenumber $\vec{k}$ is given by $-k(\hat{x}\sin(\alpha)\cos(\beta)+\hat{y}\sin(\alpha)\sin(\beta)+\hat{z}\cos(\alpha))$ where the corresponding free-space wavenumber is $k=\frac{2\pi f}{c}$ and $c$ is the speed of light. The angular spectrum $\vec{F}(\Theta)$ can be written as $\vec{F}(\Theta) = F_\alpha(\Theta)\hat{\alpha}+F_\beta(\Theta)\hat{\beta}$, where $F_\alpha(\Theta)$, $F_\beta(\Theta)$ are complex numbers and $\hat{\alpha}$, $\hat{\beta}$ are unit vectors that are orthogonal to each other and to $\vec{k}$. If the speed of the $i$-th scatterer is $v_i$, then $\vec{E}_i(t,f)$ can be represented as
\begin{eqnarray}
\label{eq: Integral representation}
\vec{E}_i(t,f) \!\!=\!\!\! \int_{0}^{2\pi}\!\!\!\!\!\int_{0}^{\pi}\!\!\! \vec{F}(\Theta)\exp(-j\vec{k}\cdotp\vec{v}_i t) \sin(\alpha)\,\mathrm{d}\alpha\,\mathrm{d}\beta,\!
\end{eqnarray}
where $z$-axis is aligned with the moving direction of scatterer $i$, as illustrated in Fig.~\ref{fig:Sphere_Scattering}, and time dependence $\exp(-j2\pi f t)$ is suppressed since it does not affect any results that will be derived later. The angular spectrum $\vec{F}(\Theta)$ could be either deterministic or random. The electric field in \eqref{eq: Integral representation} satisfies Maxwell's equations because each plane-wave component satisfies Maxwell's equations~\cite{Hill09Electromagnetic}.

Radio propagation in a building interior is in general very difficult to be analyzed because that the EM waves can be absorbed and scattered by walls, doors, windows, moving objects, etc. However, buildings and rooms can be viewed as reverberation cavities in that they exhibit internal multipath propagations. Hence, we refer to a statistical modeling instead of a deterministic one and apply the statistical theory of EM fields developed for reverberation cavities to analyze the statistical properties of $\vec{E}_{i}(t,f)$. We assume that $\vec{E}_{i}(t,f)$ is a superposition of a large number of plane waves with uniformly distributed arrival directions, polarizations, and phases, which can well capture the properties of the wave functions of reverberation cavities~\cite{Hill09Electromagnetic}. Therefore, we take $\vec{F}(\Theta)$ to be a random variable and the corresponding statistical assumptions on $\vec{F}(\Theta)$ are summarized as follows:
\begin{assumption}
\label{Assump: 1}
For $\forall \Theta$, $F_\alpha(\Theta)$ and $F_\beta(\Theta)$ are both circularly-symmetric Gaussian random variables~\cite{Tse05Fundamentals} with the same variance, and they are statistically independent.
\end{assumption}
\begin{assumption}
\label{Assump: 2}
For each dynamic scatterer, the angular spectrum components arriving from different directions are uncorrelated.
\end{assumption}
\begin{assumption}
\label{Assump: 3}
For any two dynamic scatterers $i_1,\;i_2\in\Omega_d$, $\vec{E}_{i_1}(t_1,f)$ and $\vec{E}_{i_2}(t_2,f)$ are uncorrelated, for $\forall t_1$, $t_2$.
\end{assumption}

Assumption~\ref{Assump: 1} is due to the fact that the angular spectrum is a result of many rays or bounces with random phases and thus it can be assumed that each orthogonal component of $\vec{F}(\Theta)$ tends to be Gaussian under the Central Limit Theorem. Assumption~\ref{Assump: 2} is because that the angular spectrum components corresponding to different directions have taken very different multiple scattering paths and they can thus be assumed to be uncorrelated with each other. Assumption~\ref{Assump: 3} results from the fact that the channel responses of two locations separated by at least half wavelength are statistically uncorrelated~\cite{Chen17Achieving}\cite{Zhang17time}, and the electric fields contributed by different scatterers can thus be assumed to be uncorrelated.

Under these three assumptions, $\vec{E}_i(t,f)$, $\forall i\in\Omega_d$ can be approximated as a stationary process in time. Define the temporal ACF of an electric field $\vec{E}(t,f)$ as
\begin{eqnarray}
\label{eq: defn of ACF}
\rho_{\vec{E}}(\tau,f) = \frac{\langle \vec{E}(0,f), \vec{E}(\tau,f) \rangle}{\sqrt{\langle |\vec{E}(0,f)|^2 \rangle \langle |\vec{E}(\tau,f)|^2 \rangle}},
\end{eqnarray}
where $\tau$ is the time lag, $\langle\; \rangle$ stands for the ensemble average over all realizations, $\langle \vec{X},\vec{Y} \rangle$ denotes the inner product of $\vec{X}$ and $\vec{Y}$, i.e., $\langle \vec{X},\vec{Y} \rangle\triangleq \langle \vec{X}\cdot\vec{Y}^* \rangle$ and $^*$ is the operator of complex conjugate and $\cdot$ is dot product, $|\vec{E}(t,f)|^2$ denotes the square of the absolute value of the electric field. Since $\vec{E}(t,f)$ is assumed to be a stationary process, the denominator of \eqref{eq: defn of ACF} degenerates to $E^2(f)$ which stands for the power of the electric field, i.e., $E^2(f)=\langle |\vec{E}(t,f)|^2 \rangle$, $\forall t$, and the ACF is merely a normalized counterpart of the auto-covariance function.

For the $i$-th scatterer with moving velocity $\vec{v}_i$, $\langle \vec{E}_i(0,f)\cdot \vec{E}_i^*(\tau,f) \rangle$ can be derived as~\cite{Hill09Electromagnetic}
\begin{eqnarray}
& & \langle \vec{E}_i(0,f)\cdot \vec{E}_i^*(\tau,f) \rangle \nonumber\\
&=& \int_{4\pi}\int_{4\pi} \langle\vec{F}(\Theta_1)\cdot\vec{F}(\Theta_2)\rangle \exp(j\vec{k}_2\cdotp\vec{v}_i \tau) \,\mathrm{d}\Theta_1 \,\mathrm{d}\Theta_2 \nonumber\\
&=& \frac{E_i^2(f)}{4\pi} \int_{4\pi} \exp(jk v_i \tau\cos(\alpha_2)) \mathrm{d}\Theta_2 \nonumber\\
&=& E_i^2(f) \frac{\sin(kv_i\tau)}{kv_i\tau},
\end{eqnarray}
where we define $\int_{4\pi}\triangleq \int_{0}^{2\pi}\int_{0}^{\pi}$ and $\mathrm{d}\Theta\triangleq \sin(\alpha)\,\mathrm{d}\alpha\,\mathrm{d}\beta$, and $E_i^2(f)$ is the power of $\vec{E}_i(t,f)$. With Assumption~\ref{Assump: 3}, the auto-covariance function of $\vec{E}_{Rx}(t,f)$ can be written as
\begin{eqnarray}
& & \left\langle (\vec{E}_{Rx}(0,f)-\vec{E}_s(f))\cdot (\vec{E}_{Rx}^*(\tau,f)-\vec{E}^*_s(f)) \right\rangle\nonumber\\
&=&\sum_{i\in\Omega_d} E_i^2(f) \frac{\sin(kv_i\tau)}{kv_i\tau},
\end{eqnarray}
and the corresponding ACF can thus be derived as
\begin{eqnarray}
\label{eq: ACF of E_Rx}
\rho_{\vec{E}_{Rx}}(\tau,f) \!=\! \frac{1}{\sum_{j\in\Omega_d} E_j^2(f)}\!\!\sum_{i\in\Omega_d} E_i^2(f) \frac{\sin(kv_i\tau)}{kv_i\tau}.
\end{eqnarray}
From \eqref{eq: ACF of E_Rx}, the ACF of $\vec{E}_{Rx}$ is actually a combination of the ACF of each moving scatterer weighted by their radiation power, and the moving direction of each dynamic scatterer does not play a role in the ACF. The importance of \eqref{eq: ACF of E_Rx}  lies in the fact that the speed information of the dynamic scatterers is actually embedded in the ACF of the received electric field.

\section{Theoretical Foundation of WiSpeed}
\label{sec: Theoretical Foundation of WiSpeed}

In Section~\ref{sec:Statistical theory}, we have derived the ACF of the received electric field at the Rx, which depends on the speed of the dynamic scatterers. If all or most of the dynamic scatterers move at the same speed $v$, then the right-hand side of \eqref{eq: ACF of E_Rx} would degenerate to $\rho_{\vec{E}_{Rx}}(\tau,f) = \frac{\sin(kv\tau)}{kv\tau}$, and it becomes very simple to estimate the common speed from the ACF. However, it is not easy to directly measure the electric field at the Rx and analyze its ACF. Instead, the power of the electric field can be viewed equivalent to the power of the channel response that can be measured by commercial WiFi devices. In this section, we will discuss the principle of WiSpeed that utilizes the ACF of the CSI power response for speed estimation.

Without loss of generality, we use the channel response of OFDM-based WiFi systems as an example. Let $X(t,f)$ and $Y(t,f)$ be the transmitted and received signals over a subcarrier with frequency $f$ at time $t$. Then, the least-square estimator of the CSI for the subcarrier with frequency $f$ measured at time $t$ is $H(t,f) = \frac{Y(t,f)}{X(t,f)}$~\cite{Chiueh12Baseband}. We define the power response $G(t,f)$ as the square of the magnitude of CSI, which takes the form
\begin{eqnarray}
\label{eq: Power response}
G(t,f) \triangleq |H(t,f)|^2 = \|\vec{E}_{Rx}(t,f)\|^2 + \varepsilon(t,f),
\end{eqnarray}
where $\|\vec{E}\|^2$ denotes the total power of $\vec{E}$, and $\varepsilon(t,f)$ is assumed to be an additive noise due to the imperfect measurement of CSI.

\begin{figure}[tb]
    \centering
    \begin{subfigure}[b]{0.24\textwidth}
        \includegraphics[width=\textwidth]{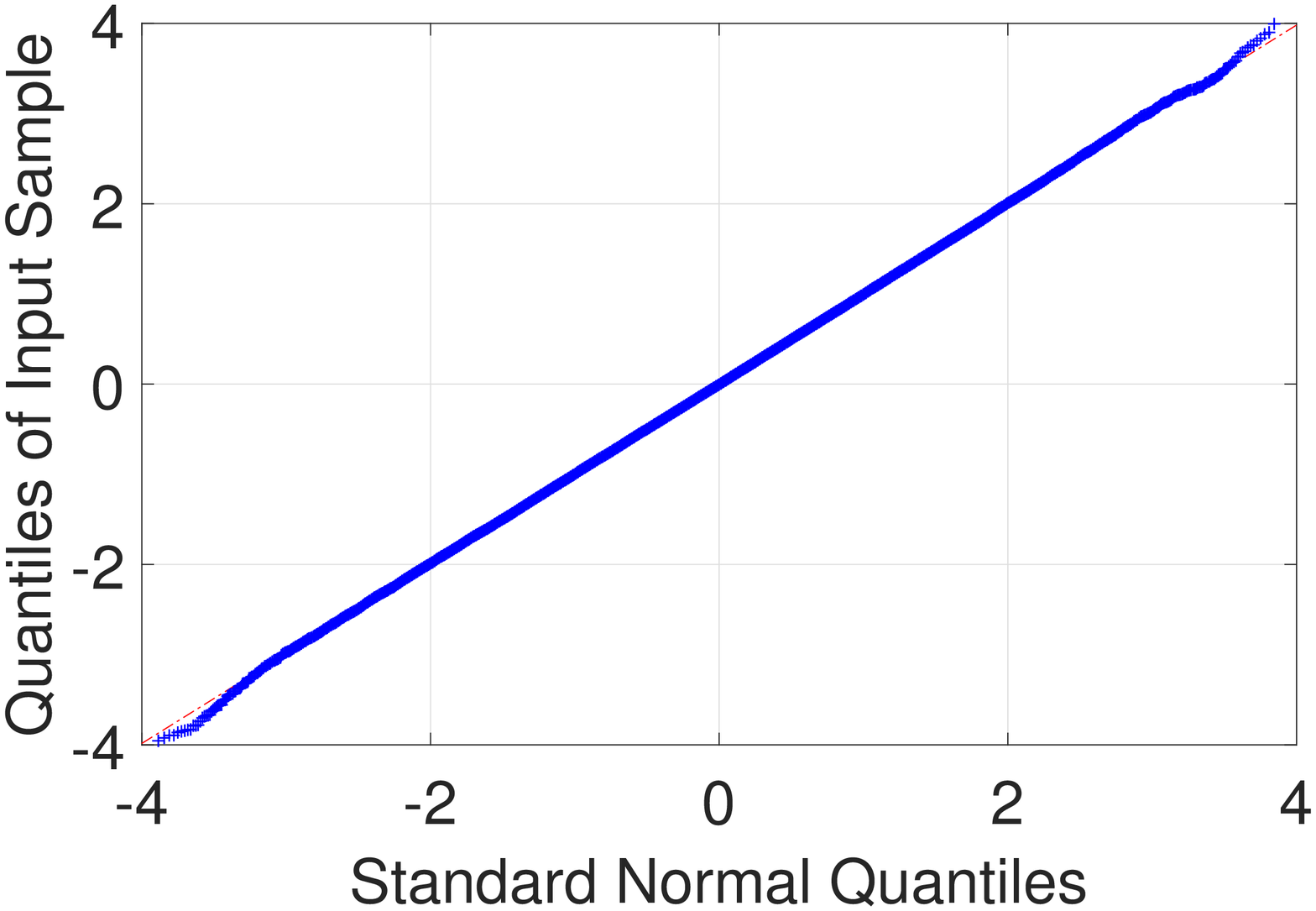}
        \caption{\label{fig:QQplot} Q-Q plot of noise.}
    \end{subfigure}
    ~
    \begin{subfigure}[b]{0.22\textwidth}
        \includegraphics[width=\textwidth]{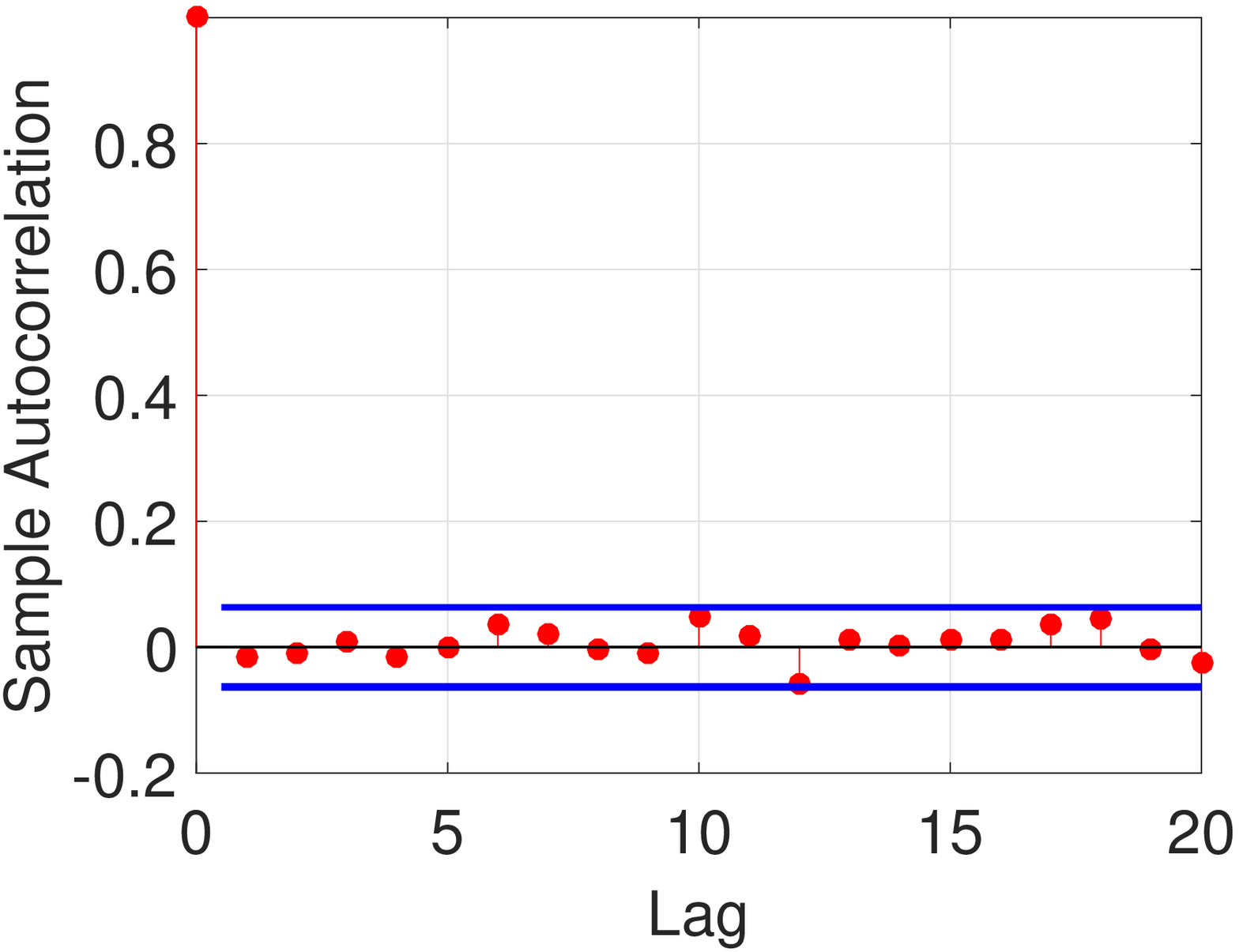}
        \caption{\label{fig:SampleACF_noise} Sample ACF of noise.}
    \end{subfigure}
    \caption{\label{fig:Noise_Modelling} The Q-Q plot and sample ACF of a typical CSI power response.}
\end{figure}

The noise $\varepsilon(t,f)$ can be assumed to follow a normal distribution. To prove this, we collect a set of one-hour CSI data in a static indoor environment with the channel sampling rate $F_s=30\,$Hz. The Q-Q plot of the normalized $G(t,f)$ and standard normal distribution for a given subcarrier is shown in Fig.~\ref{fig:QQplot}, which shows that the distribution of the noise is very close to a normal distribution. To verify the whiteness of the noise, we also study the ACF of $G(t,f)$ that can be defined as~\cite{Shumway10Time} $\rho_G(\tau,f) = \frac{\gamma_G(\tau,f)}{\gamma_G(0,f)}$, where $\gamma_G(\tau,f)$ denotes the auto-covariance function, i.e., $\gamma_G(\tau,f)\triangleq\mathrm{cov}(G(t,f),G(t-\tau,f))$. In practice, sample auto-covariance function $\hat{\gamma}_G(\tau,f)$ is used instead. If $\varepsilon(t,f)$ is white noise, the sample ACF $\hat{\rho}_G(\tau,f)$, for $\forall \tau\neq 0$, can be approximated by a normal random variable with zero mean and standard deviation $\sigma_{\hat{\rho_G}(\tau,f)}=\frac{1}{\sqrt{T}}$. Fig.~\ref{fig:SampleACF_noise} shows the sample ACF of $G(t,f)$ when $2000$ samples on the first subcarrier are used. As we can see from the figure, all the taps of the sample ACF are within the interval of $\pm 2\sigma_{\hat{\rho_G}(\tau,f)}$, and thus, it can be assumed that $\varepsilon(t,f)$ is an additive white Gaussian noise, i.e., $\varepsilon(t,f)\sim\mathcal{N}(0,\sigma^2(f))$.

In the previous analysis in Section~\ref{sec:Statistical theory}, we assume that the Tx transmits continuous EM waves, but in practice the transmission time is limited. For example, in IEEE 802.11n WiFi systems operated in $5\,$GHz frequency band with $40\,$MHz bandwidth channels, a standard WiFi symbol is $4\,\mu \mathrm{s}$, composed of a $3.2\,\mu \mathrm{s}$ useful symbol duration and a $0.8\,\mu \mathrm{s}$ guard interval. According to \cite{Nee97Delay}, for most office buildings, the delay spread is within the range of $40$ to $70\,$ns, which is much smaller than the duration of a standard WiFi symbol. Therefore, we can assume continuous waves are transmitted in WiFi systems.

Based on the above assumptions and \eqref{eq: E_RX approximated decomposition}, \eqref{eq: Power response} can be approximated as
\begin{eqnarray}
& & G(t,f)\approx\| \vec{E}_s(f) + \sum_{i\in\Omega_d}\vec{E}_i(t,f) \|^2 +\varepsilon(t,f)   \nonumber \\
	   &=&\!\!\!\left\| \sum_{u\in\{x,y,z\}}\!\!\left(E_{su}(f)\hat{u} \!+\! \sum_{i\in\Omega_d}E_{iu}(t,f)\hat{u}\!\right) \!\right\|^2 \!+\!\varepsilon(t,f) \nonumber \\
       &=&\!\!\!\sum_{u\in\{x,y,z\}}\left|E_{su}(f) + \sum_{i\in\Omega_d}E_{iu}(t,f)\right|^2+\varepsilon(t,f)\nonumber \\
       &=&\!\!\!\sum_{u\in\{x,y,z\}}\!\!\!\bigg(|E_{su}(f)|^2 \!+\! 2\mathrm{Re}\left\{\!E^*_{su}(f)\sum_{i\in\Omega_d}E_{iu}(t,f)\!\!\right\}\nonumber\\
       & & +\left|\sum_{i\in\Omega_d}E_{iu}(t,f)\right|^2 \bigg)+\varepsilon(t,f),
\end{eqnarray}
where $\hat{x}$, $\hat{y}$ and $\hat{z}$ are unit vectors orthogonal to each other as shown in Fig.~\ref{fig:Sphere_Scattering}, $\mathrm{Re}\{\cdot\}$ denotes the operation of taking the real part of a complex number, and $E_{iu}$ denotes the component of $\vec{E}_i$ in the $u$-axis direction, for $\forall u\in\{x,y,z\}$. Then, the auto-covariance function of $G(t,f)$ can be derived as
\begin{eqnarray}
\label{eq: cov of G}
 & &  \gamma_G(\tau,f)=\mathrm{cov}\left(G(t,f),G(t-\tau,f)\right) \nonumber \\
 \!\! \!\!\!\!&\approx&\!\!\!\!\!\!\! \sum_{u\in\{x,y,z\}}\!\!\!\! \Bigg(\!\!2|E_{su}(f)|^2\sum_{i\in\Omega_d} \mathrm{cov}(E_{iu}(t,f)\!,\!E_{iu}(t\!-\!\tau,f)) \nonumber \\
  & & +\!\! \sum\limits_{\substack{i_1,i_2\in\Omega_d\\i_1\geq i_2}}\!\!\!\!\mathrm{cov}(E_{i_1 u}(t,f), E_{i_1 u}(t-\tau,f))\cdot \nonumber \\
 \!\!\!& &\!\!\!\! \mathrm{cov}(E_{i_2 u}(t,f),E_{i_2 u}(t-\tau,f))\!\!\Bigg) \!\!+\!\! \delta(\tau)\sigma^2(f),
\end{eqnarray}
where Assumptions~\ref{Assump: 1}-\ref{Assump: 3} and \eqref{eq: Integral representation} are applied to simplify the expression and the detailed derivations can be found in Appendix \ref{App:1}.

\begin{figure}[tb]
    \centering
    \begin{subfigure}[b]{0.23\textwidth}
        \includegraphics[width=\textwidth]{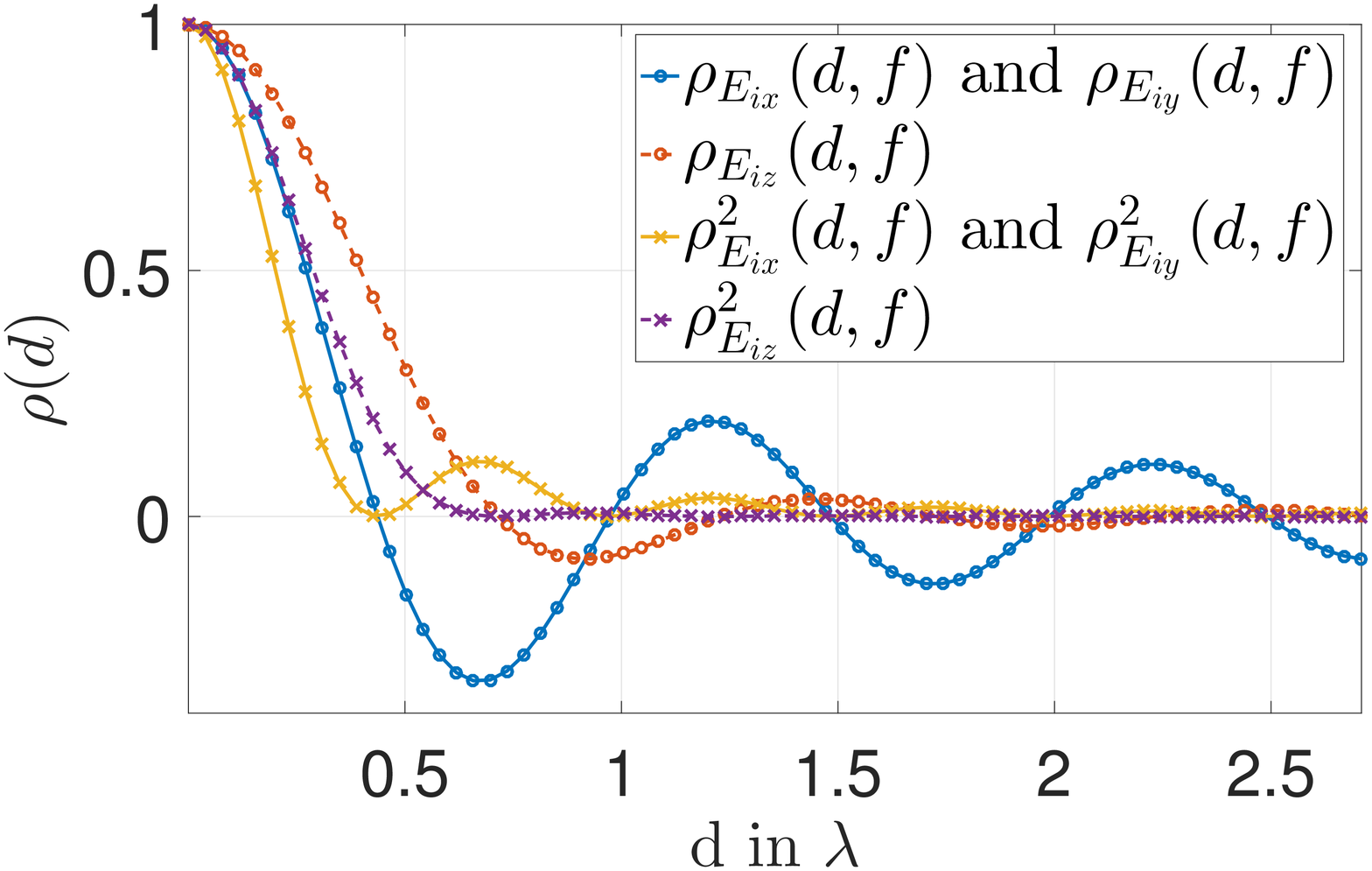}
        \caption{\label{fig:TheoreticalSpatialACF} Theoretical spatial ACFs.}
    \end{subfigure}
    ~
    \begin{subfigure}[b]{0.23\textwidth}
        \includegraphics[width=\textwidth]{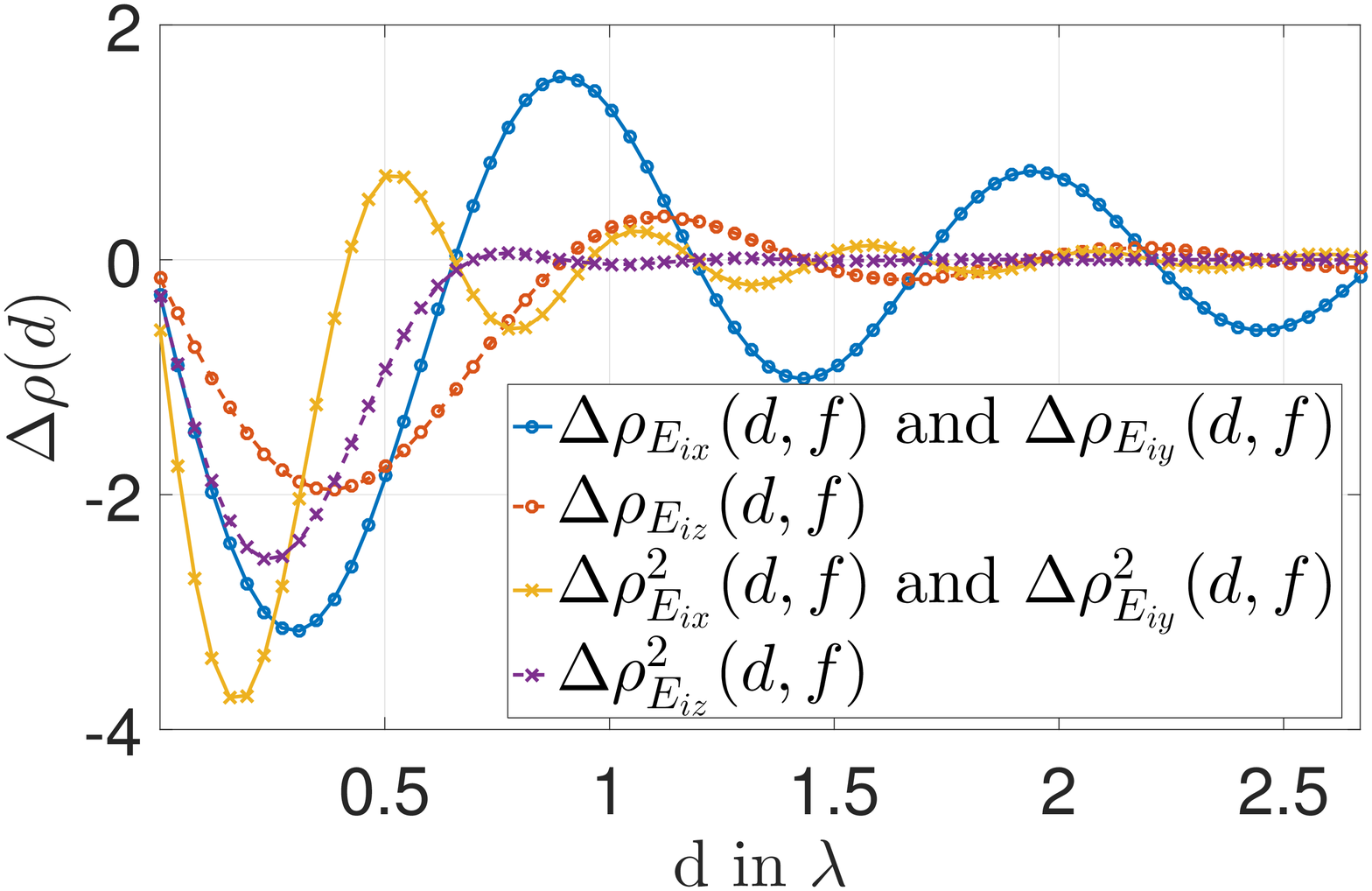}
        \caption{\label{fig:DiffTheoreticalSpatialACF} Diff. of spatial ACFs.}
    \end{subfigure}
    \caption{Theoretical spatial ACF for different orthogonal components of EM waves.}
\end{figure}

According to the relation between the auto-covariance and autocorrelation, $\gamma_G(\tau,f)$ can be rewritten in the forms of ACFs of each scatterer as
\begin{eqnarray}
\label{eq: gamma_G(tau)}
\!\!\!\!& & \gamma_G(\tau,f) \!\!\approx\!\!\!\! \sum_{u\in\{x,y,z\}} \!\!\!\Bigg(\!\!\sum_{i\in\Omega_d}\frac{2|E_{su}(f)|^2 E^2_{i}(f)}{3}\rho_{E_{iu}}(\tau,f) \nonumber \\
\!\!\!\!& & \!\!\!\!+\!\!\!\!\!\! \sum\limits_{\substack{i_1, i_2\in\Omega_d\\i_1\geq i_2}}\!\!\!\!\!\! \frac{E^2_{i_1}(\!f\!)E^2_{i_2}(\!f\!)}{9}\rho_{E_{i_1 u}}(\!\tau\!,\!f\!) \rho_{E_{i_2 u}}(\!\tau\!,\!f\!)\!\!\Bigg)\!\!+\!\! \delta(\!\tau\!)\sigma^2(\!f\!),
\end{eqnarray}
where the right-hand side is obtained by using the relation $E^2_{iu}(f) = \frac{E^2_i(f)}{3}$, $\forall u\in\{x,y,z\}$, $\forall i\in\Omega_d$~\cite{Hill09Electromagnetic}. The corresponding ACF $\rho_G(\tau,f)$ of $G(t,f)$ is thus obtained by $\rho_G(\tau,f) = \frac{\gamma_G(\tau,f)}{\gamma_G(0,f)}$, where $\gamma_G(\tau,0)$ can be obtained by plugging $\rho_{E_{iu}}(0,f)=1$ into \eqref{eq: gamma_G(tau)}. When the moving directions of all the dynamic scatterers are approximately the same, then we can choose $z$-axis aligned with the common moving direction. Then, the closed forms of $\rho_{E_{iu}}(\tau,f)$, $\forall u\in\{x,y,z\}$, are derived under Assumptions~\ref{Assump: 1}-\ref{Assump: 2}~\cite{Hill09Electromagnetic}, i.e., for $\forall i\in\Omega_d$,
\begin{eqnarray}
\!\!\!\!\!\!& &\rho_{E_{ix}}(\tau,f) = \rho_{E_{iy}}(\tau,f) \nonumber\\
\!\!\!\!\!\!&=&\!\!\! \frac{3}{2}\!\bigg[\!\!\frac{\sin(\!kv_i\tau\!)}{kv_i\tau}\!\!-\!\!\frac{1}{(kv_i\tau)^2}\!\!\left(\!\!\frac{\sin(kv_i\tau)}{kv_i\tau}\!-\!\cos(kv_i\tau)\!\!\right)\!\!\!\bigg], \\
\!\!\!\!\!\!& &\!\!\!\!\rho_{E_{iz}}(\tau,f) \!=\!  \frac{3}{(kv_i\tau)^2}\!\left[\frac{\sin(kv_i\tau)}{kv_i\tau}\!-\!\cos(kv_i\tau)\right].
\end{eqnarray}
The theoretical spatial ACFs are shown in Fig.~\ref{fig:TheoreticalSpatialACF} where $d\triangleq v_i\tau$. As we can see from Fig.~\ref{fig:TheoreticalSpatialACF}, the magnitudes of all the ACFs decay with oscillations as the distance $d$ increases.

For a WiFi system with a bandwidth of $40\,$MHz and a carrier frequency of $5.805\,$GHz, the difference in the wavenumber $k$ of each subcarrier can be neglected, e.g., $k_{\mathrm{max}}=122.00$ and $k_{\mathrm{min}}=121.16$. Then, we can assume $\rho(\tau,f)\approx\rho(\tau)$, $\forall f$. Thus, we can improve the sample ACF by averaging across all subcarriers, i.e., $\hat{\rho}_G(\tau) \triangleq \frac{1}{F} \sum_{f\in\mathcal{F}} \hat{\rho}_G(\tau,f)$, where $\mathcal{F}$ denotes the set of all the available subcarriers and $F$ is the total number of subcarriers. When all the dynamic scatterers have the same speed, i.e., $v_i=v$ for $\forall i\in\Omega_d$, which is the case for monitoring the motion for a single human subject, by defining the substitutions $E^2_{su}\triangleq \frac{2}{F} \sum_{f\in\mathcal{F}} |E_{su}(f)|^2$, $E^2_d\triangleq \frac{1}{3F} \sum_{i\in\Omega_d} \sum_{f\in\mathcal{F}} E^2_i(f)$, $\hat{\rho}_G(\tau)$ can be further approximated as (for $\tau\neq 0$)
\begin{eqnarray}
\label{eq: approx of rho_G}
\hat{\rho}_G(\tau) \approx C \sum_{u\in\{x,y,z\}} \bigg( E^2_d \hat{\rho}^2_{E_{iu}}(\tau) + E^2_{su} \hat{\rho}_{E_{iu}}(\tau) \bigg),
\end{eqnarray}
where $C$ is a scaling factor and the variance of each subcarrier is assumed to be close to each other.

From \eqref{eq: approx of rho_G}, we observe that $\rho_G(\tau)$ is a weighted combination of $\rho_{E_{iu}}(\tau)$ and $\rho^2_{E_{iu}}(\tau)$, $\forall u\in\{x,y,z\}$. The left-hand side of \eqref{eq: approx of rho_G} can be estimated from CSI and the speed is embedded in each term on the right-hand side. If we can separate one term from the others on the right-hand side of \eqref{eq: approx of rho_G}, then the speed can be estimated.

Taking the differential of all the theoretical spatial ACFs as shown in Fig.~\ref{fig:DiffTheoreticalSpatialACF} where we use the notation $\Delta\rho(\tau)$ to denote $\frac{\mathrm{d}\rho(\tau)}{\mathrm{d}\tau}$, we find that although the ACFs of different components of the received EM waves are superimposed, the first local peak of $\Delta\rho^2_{E_{iu}}(\tau)$, $\forall u\in\{x,y\}$, happens to be the first local peak of $\Delta\rho_{G}(\tau)$ as well. Therefore, the component $\rho^2_{E_{iu}}(\tau)$ can be recognized from $\rho_{G}(\tau)$, and the speed information can thus be obtained by localizing the first local peak of $\Delta\hat{\rho}_G(\tau)$, which is the most important feature that WiSpeed extracts from the noisy CSI measurements.

To verify \eqref{eq: approx of rho_G}, we build a prototype of WiSpeed with commercial WiFi devices. The configurations of the prototype are summarized as follows: both WiFi devices operate on WLAN channel $161$ with a center frequency of $f_{c}=5.805\,$GHz, and the bandwidth is $40\,$MHz; the Tx is equipped with a commercial WiFi chip and two omnidirectional antennas, while the Rx is equipped with three omnidirectional antennas and uses Intel Ultimate N WiFi Link 5300 with modified firmware and driver~\cite{Halperin11tool}. The Tx sends sounding frames with a channel sampling rate $F_{s}$ of $1500\,$Hz, and CSI is obtained at the Rx. The transmission power is configured as $20\,$dBm.

\begin{figure}[tb]
    \centering
    \includegraphics[width=0.4\textwidth]{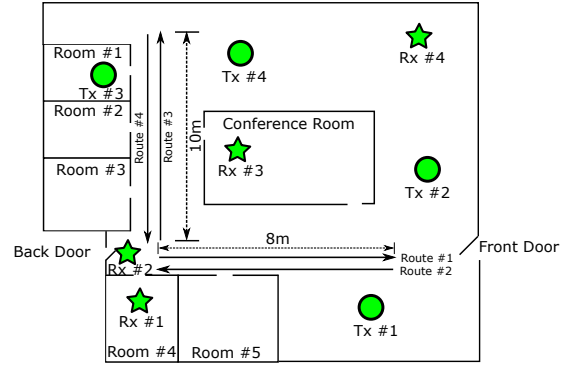}
    \caption{\label{fig:Experiment_Map} Experimental settings in a typical office environment with different Tx/Rx locations and walking routes.}
\end{figure}

All experiments in this paper are conducted in a typical indoor office environment as shown in Fig.~\ref{fig:Experiment_Map}. In each experiment, the LOS path between the Tx and the Rx is blocked by at least one wall, resulting in a severe NLOS condition. More specifically, we investigate two cases:
\begin{enumerate}
\item \textbf{The Tx is in motion and the Rx remains static:} The Tx is attached to a cart and the Rx is placed at Location Rx \#$1$ as shown in Fig.~\ref{fig:Experiment_Map}. The cart is pushed forward at an almost constant speed along Route \#$1$ marked in Fig.~\ref{fig:Experiment_Map} from $t=3.7\,$s to $t=14.3\,$s.
\item \textbf{Both the Tx and the Rx remain static and a person passes by:} the Tx and Rx are placed at Location Tx \#$1$ and Rx \#$1$ respectively. A person walks along Route \#$1$ at a speed similar to Case (1) from $t=4.9\,$s to $t=16.2\,$s.
\end{enumerate}

Since the theoretical approximations are only valid under the short duration assumption, we set the maximum time lag $\tau$ as $0.2\,$s. In both cases, we compute the sample ACF $\hat{\rho}_G(\tau)$ every $0.05\,$s.

\begin{figure}[tb]
    \centering
    \begin{subfigure}[b]{0.23\textwidth}
        \includegraphics[width=\textwidth]{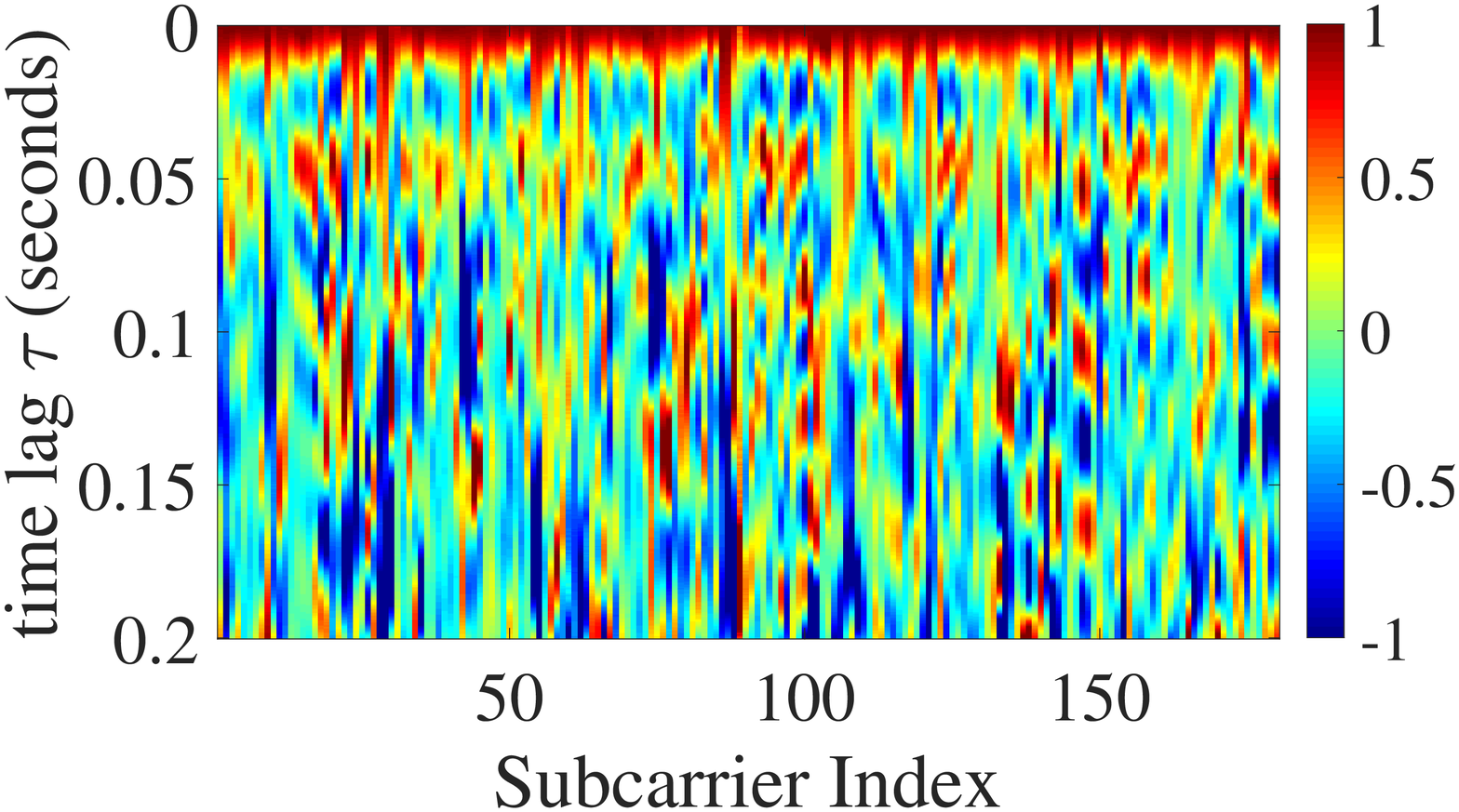}
        \caption{\label{fig:ACF_ActiveSource_Different_Subcarriers} ACF measured by different subcarriers for a moving Tx.}
    \end{subfigure}
    ~
    \begin{subfigure}[b]{0.23\textwidth}
        \includegraphics[width=\textwidth]{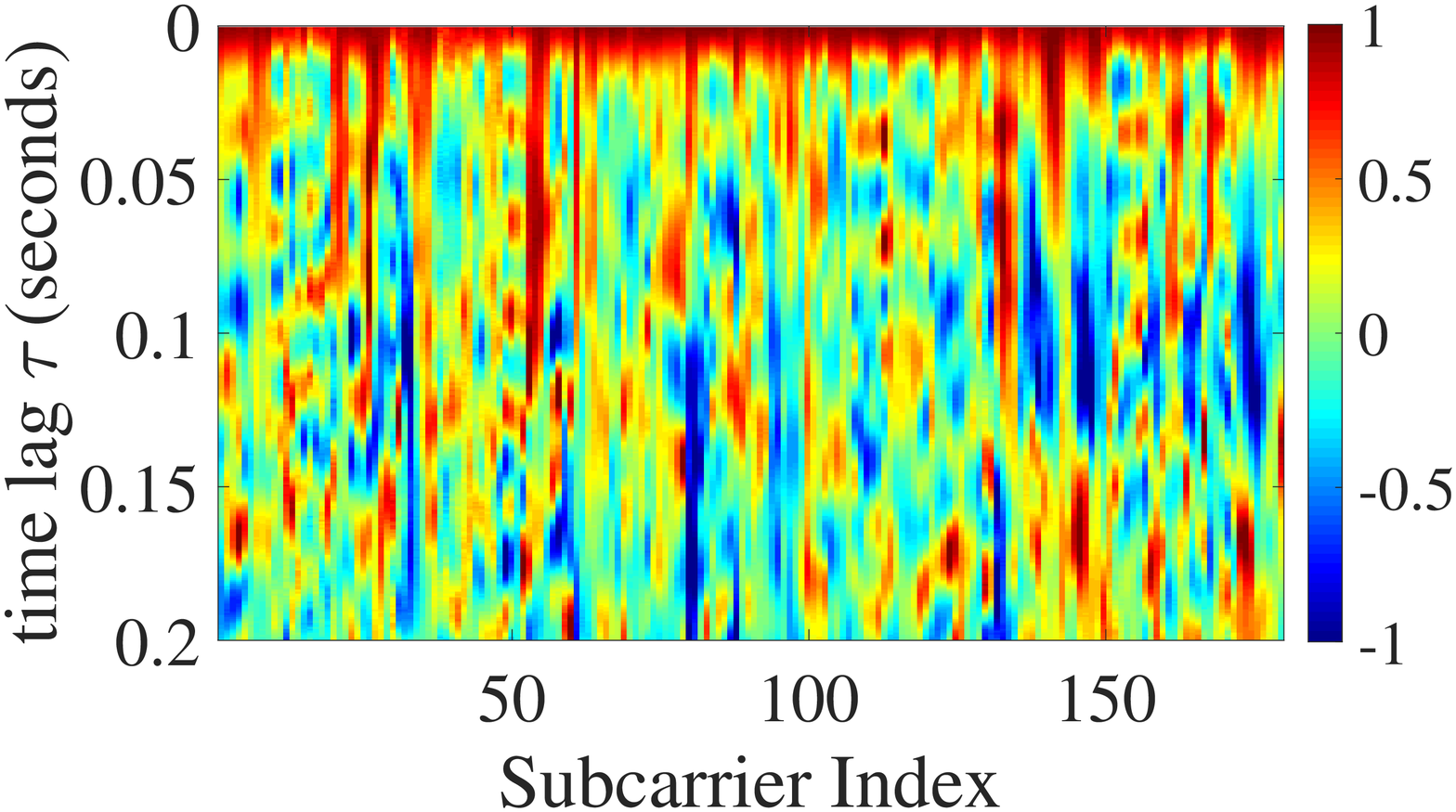}
        \caption{\label{fig:ACF_PassiveSource_Different_Subcarriers} ACF measured by different subcarriers for a walking human.}
    \end{subfigure}
    ~
    \begin{subfigure}[b]{0.23\textwidth}
        \includegraphics[width=\textwidth]{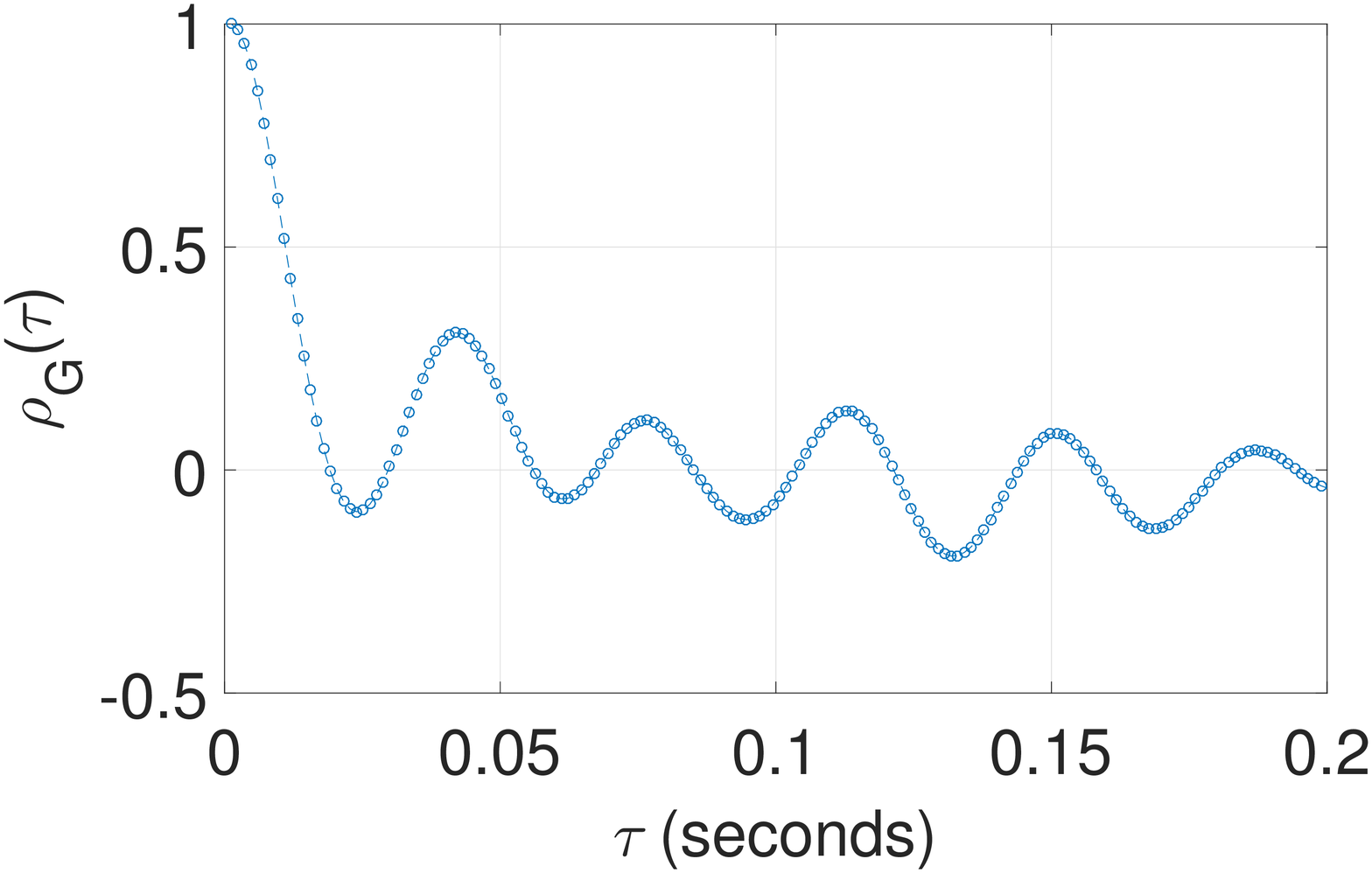}
        \caption{\label{fig:ACF_ActiveSource_oneshot} Snapshot of ACF for a moving Tx.}
    \end{subfigure}
  	~
    \begin{subfigure}[b]{0.23\textwidth}
        \includegraphics[width=\textwidth]{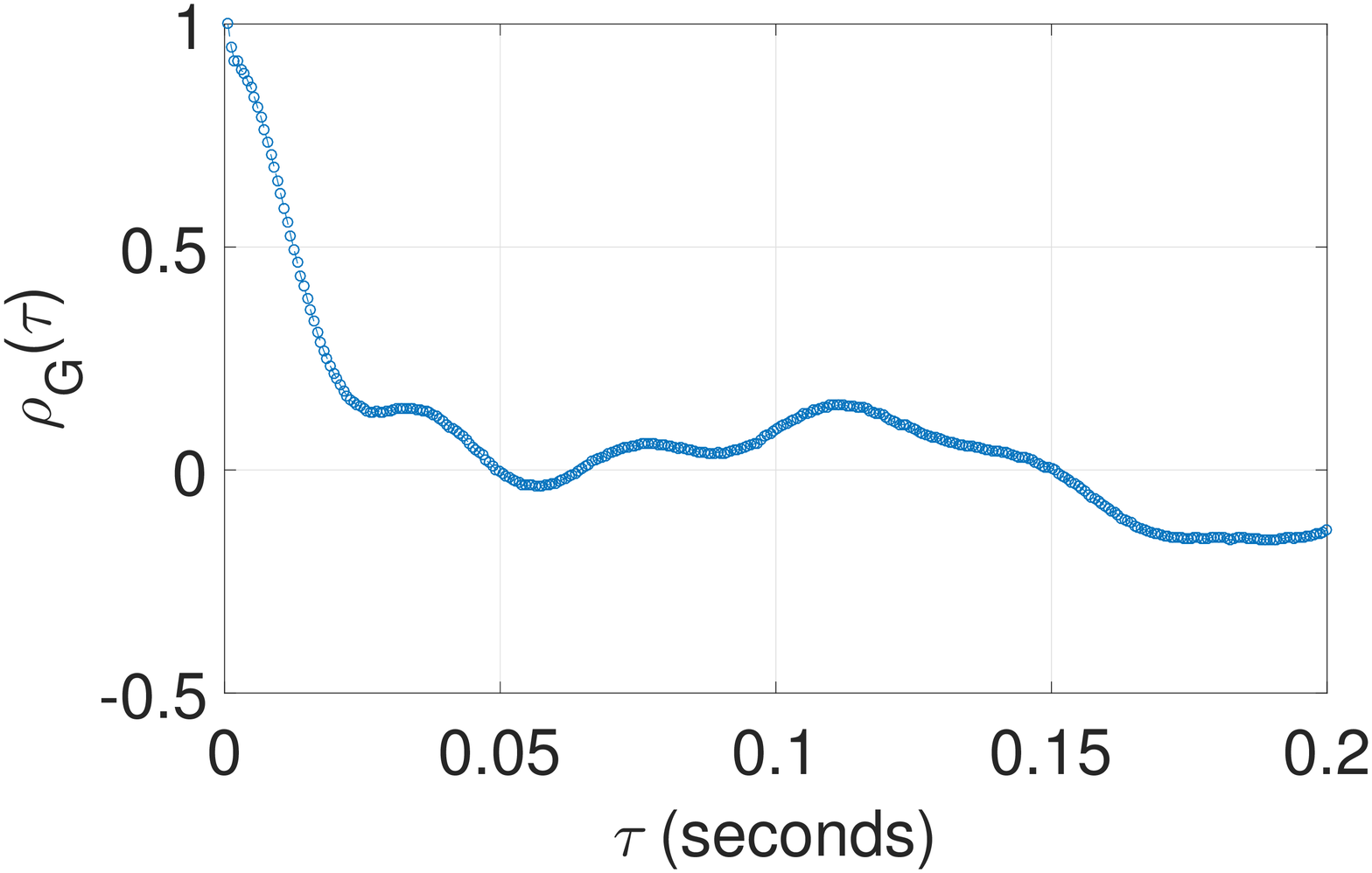}
        \caption{\label{fig:ACF_PassiveSource_oneshot} Snapshot of ACF for a walking human.}
    \end{subfigure}
    ~
    \begin{subfigure}[b]{0.23\textwidth}
        \includegraphics[width=\textwidth]{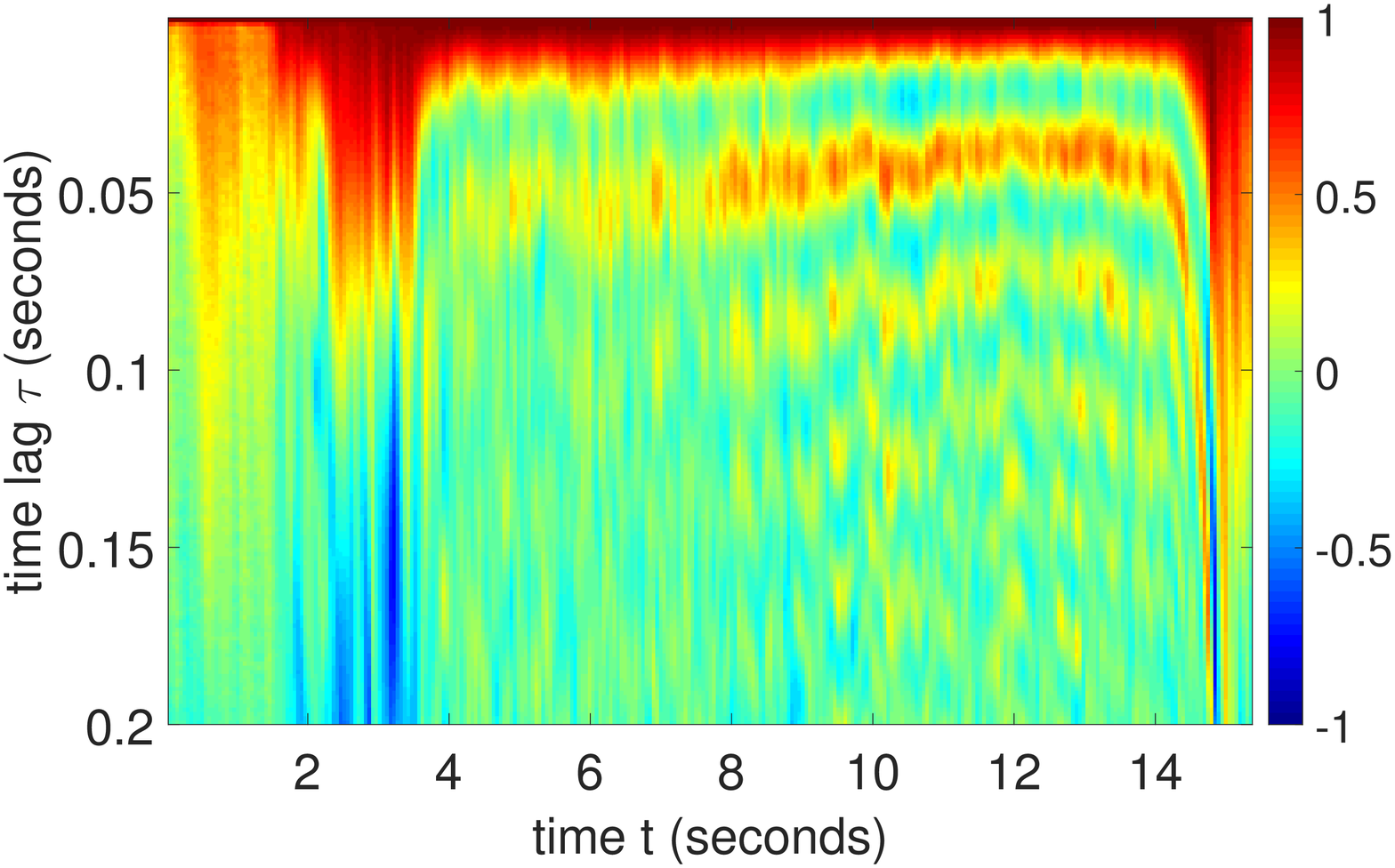}
        \caption{\label{fig:ACF_ActiveSource} ACF matrix for a moving Tx.}
    \end{subfigure}
    ~
    \begin{subfigure}[b]{0.23\textwidth}
        \includegraphics[width=\textwidth]{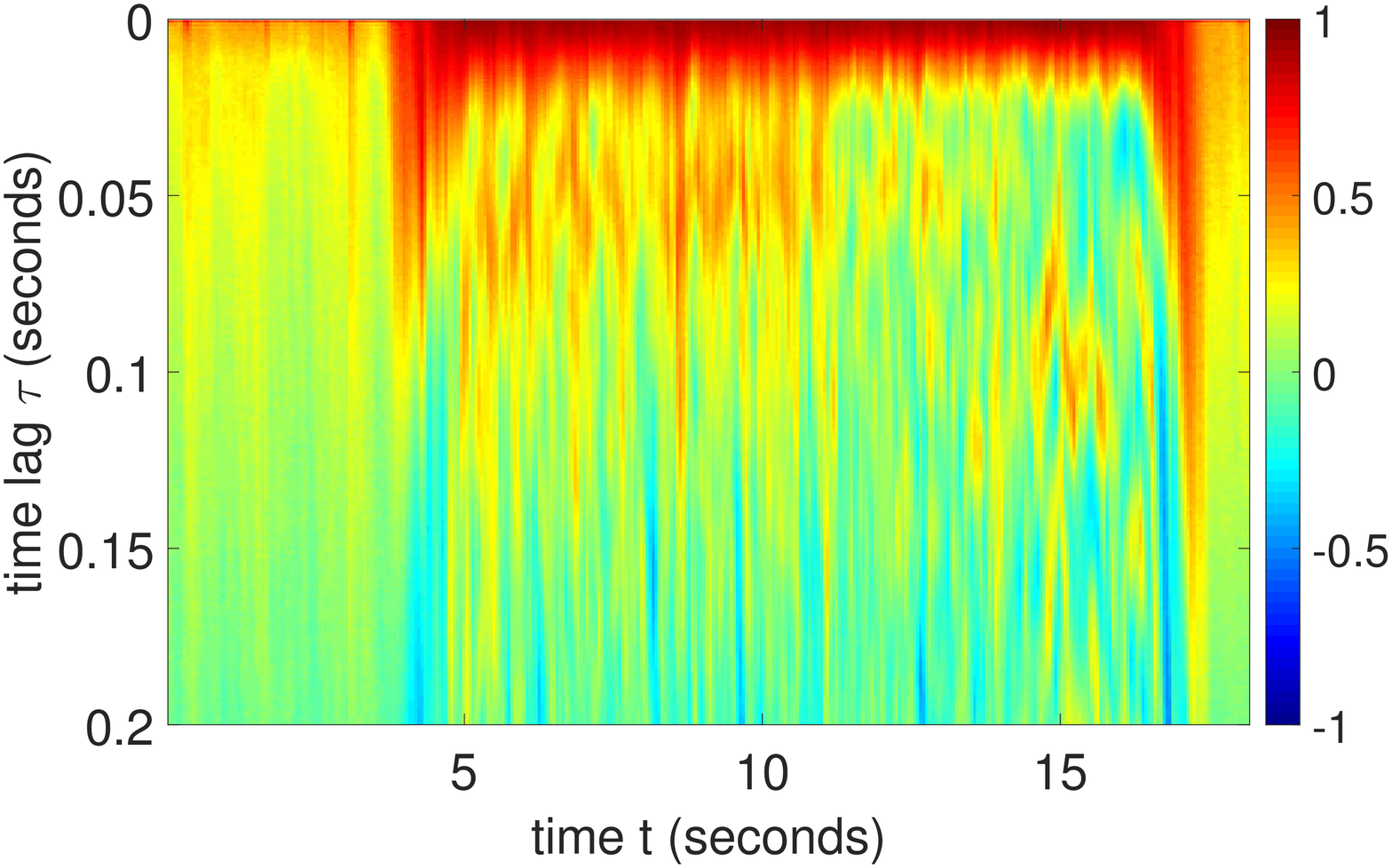}
        \caption{\label{fig:ACF_PassiveSource} ACF matrix for a walking human.}
    \end{subfigure}
    \caption{\label{fig: ACF for two cases} ACFs for the two scenarios.}
\end{figure}

Fig.~\ref{fig: ACF for two cases} demonstrates the sample ACFs for the two cases. In particular, Fig.~\ref{fig:ACF_ActiveSource_Different_Subcarriers} visualizes the sample ACF corresponding to a snapshot of Fig.~\ref{fig:ACF_ActiveSource} for different subcarriers given a fixed time $t$ with the time lag $\tau\in[0,0.2s]$, and Fig.~\ref{fig:ACF_ActiveSource_oneshot} shows the average ACF $\hat{\rho}_G(\tau)$, which is much less noisy compared with individual $\hat{\rho}_G(\tau, f)$. In this case, the Tx can be regarded as a moving scatterer with a dominant radiation power compared with the other scatterers, giving rise to the dominance of $E^2_d \rho^2_{E_{iu}}(\tau)$, $u\in\{x,y,z\}$ over the other components in \eqref{eq: approx of rho_G}. Additionally, $\rho^2_{E_{iz}}(\tau)$ decays much faster than $\rho^2_{E_{ix}}(\tau)$ and $\rho^2_{E_{iy}}(\tau)$, and $\rho^2_{E_{ix}}(\tau) = \rho^2_{E_{iy}}(\tau)$. Thus, a similar pattern between $\hat{\rho}_G(\tau)$ and $\rho^2_{E_{ix}}(\tau)$ ($\rho^2_{E_{iy}}(\tau)$) can be observed with a common and dominant component $\frac{\sin^2(kv\tau)}{(kv\tau)^2}$, where $v$ is the speed of the cart and the person. The experimental result illustrated in Fig.~\ref{fig:ACF_ActiveSource_oneshot} matches well with the theoretical analysis.

Similarly, for Case (2), Fig.~\ref{fig:ACF_PassiveSource_Different_Subcarriers} shows the sample ACF $\hat{\rho}_G(\tau,f)$ for different subcarriers and Fig.~\ref{fig:ACF_PassiveSource_oneshot} shows the average sample ACF $\hat{\rho}_G(\tau)$, which is a snapshot of Fig.~\ref{fig:ACF_PassiveSource} given a fixed time $t$ with the time lag $\tau=[0,0.2s]$. Clearly, the pattern of the component $\rho^2_{E_{iu}}(\tau)$, $u\in\{x,y\}$, in the sample ACF is much less pronounced than Case (1) shown in Fig.~\ref{fig:ACF_ActiveSource_oneshot} and Fig.~\ref{fig:ACF_ActiveSource}. This can be justified by the fact that the radiation power $E^2_d$ is much smaller than that in Case (1), as the set of dynamic scatterers only consists of different parts of a human body in mobility. Consequently, the shape of $\hat{\rho}_G(\tau)$ resembles more closely to $\rho_{E_{iu}}(\tau)$, $\forall u\in\{x,y,z\}$ with a dominant component $\frac{\sin(kv\tau)}{kv\tau}$. Moreover, from Fig.~\ref{fig:ACF_PassiveSource_oneshot}, we can observe a superposition of $\frac{\sin(kv\tau)}{kv\tau}$ and $\frac{\sin^2(kv\tau)}{(kv\tau)^2}$ and the weight of $\frac{\sin(kv\tau)}{kv\tau}$ is larger than that of $\frac{\sin^2(kv\tau)}{(kv\tau)^2}$. We also observe that the embedded component $\frac{\sin^2(kv\tau)}{(kv\tau)^2}$ has a similar pattern compared with Case (1) since the moving speeds in the two experiments are similar to each other.

\section{Key Components of WiSpeed}
\label{sec: Key components of WiSpeed}

Based on the theoretical results derived in Section~\ref{sec: Theoretical Foundation of WiSpeed}, we propose WiSpeed, which integrates three modules: moving speed estimator, acceleration estimator, and gait cycle estimator. The moving speed estimator is the core module of WiSpeed, while the other two extract useful features from the moving speed estimator to detect falling down and to estimate the gait cycle of a walking person.

\subsection{Moving Speed Estimator}
\label{subsec:movingspeed}

WiSpeed estimates the moving speed of the subject by calculating the sample ACF $\Delta\hat{\rho}_G(\tau)$ from CSI measurements, localizing the first local peak of $\Delta\hat{\rho}_G(\tau)$, and mapping the peak location to the speed estimation. Since in general, the sample ACF $\Delta\hat{\rho}_G(\tau)$ is noisy as can be seen in Fig.~\ref{fig:ACF_ActiveSource} and Fig~\ref{fig:ACF_PassiveSource}, we develop a novel robust local peak identification algorithm based on the idea of local regression~\cite{Cleveland79} to reliably detect the location of the first local peak of $\Delta\hat{\rho}_G(\tau)$.

For notational convenience, write the discrete signal for local peak detection as $y[n]$, and our goal is to identify the local peaks in $y[n]$. First of all, we apply a moving window with length $2L+1$ to $y[n]$, where $L$ is chosen to be comparable with the width of the desired local peaks. Then, for each window with its center located at $n$, we verify if there exists any potential local peak within the window by performing a linear regression and a quadratic regression to the data inside the window, separately. Let $\mathrm{SSE}$ denote the sum of squared errors for the quadratic regression and $\mathrm{SSE}_r$ denote that for the linear regression. If there is no local peak within the given window, the ratio $\alpha[n]\triangleq\frac{(\mathrm{SSE}_r-\mathrm{SSE})/(3-2)}{\mathrm{SSE/(2L+1-3)}}$ can be interpreted as a measure of the likelihood of the presence of a peak within the window, and has a central F-distribution with $1$ and $2(L-1)$ degrees of freedom, under certain assumptions~\cite{Scheffe99ANOVA}. We choose a potential window with the center point $n$ only when $\alpha[n]$ is larger than a preset threshold $\eta$, which is determined by the desired probability of finding a false peak, and $\alpha[n]$ should also be larger than its neighborhoods $\alpha[n-L]$,...,$\alpha[n+L]$. When $L$ is small enough and there exists only one local peak within the window, the location of the local peak can be directly obtained from the fitted quadratic curve.

\begin{figure}[tb]
    \centering
    \begin{subfigure}[b]{0.23\textwidth}
        \includegraphics[width=\textwidth]{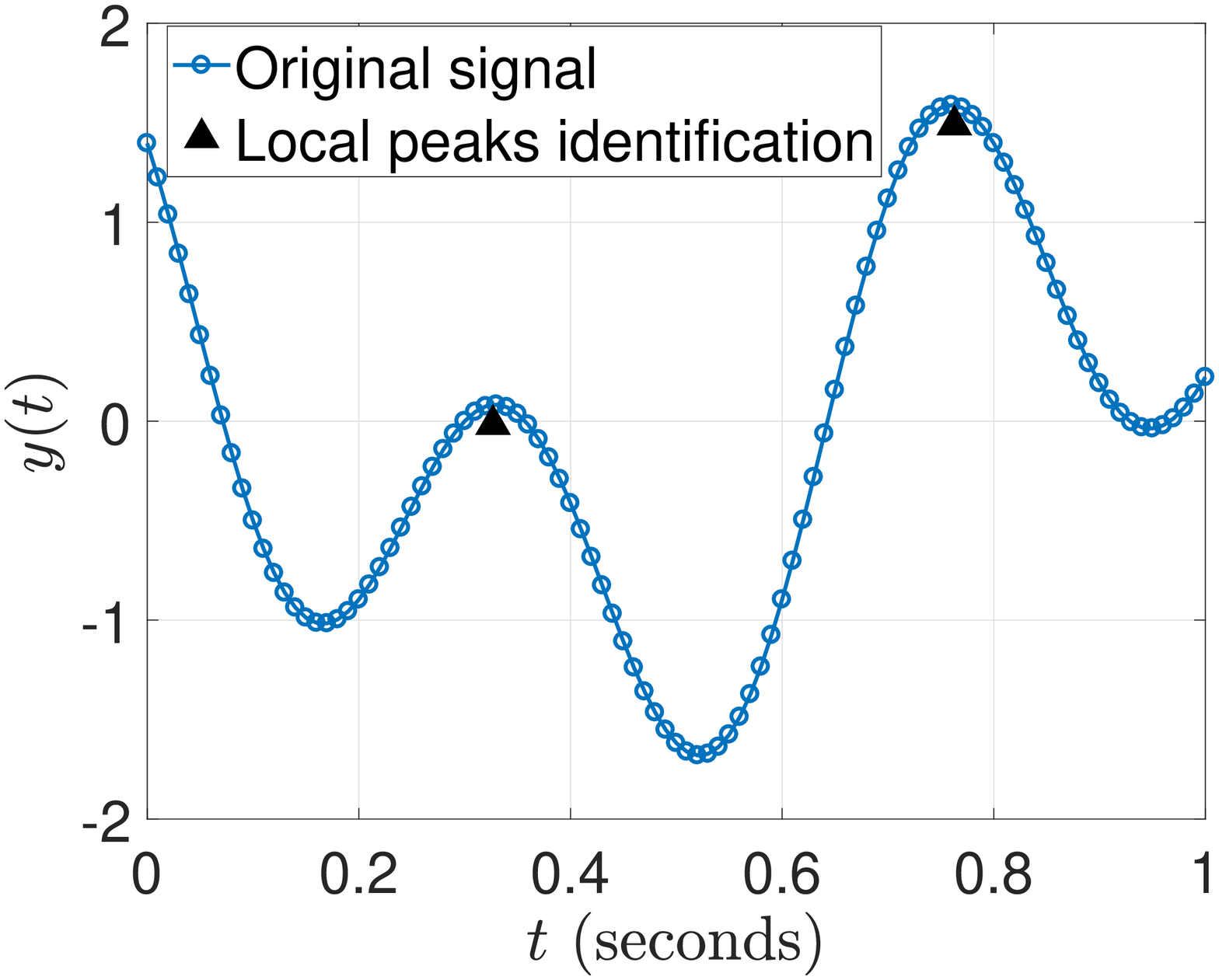}
        \caption{\label{fig:Peak_Detection_OriginalSignal} Original signal and its estimated local peaks.}
    \end{subfigure}
    ~
    \begin{subfigure}[b]{0.23\textwidth}
        \includegraphics[width=\textwidth]{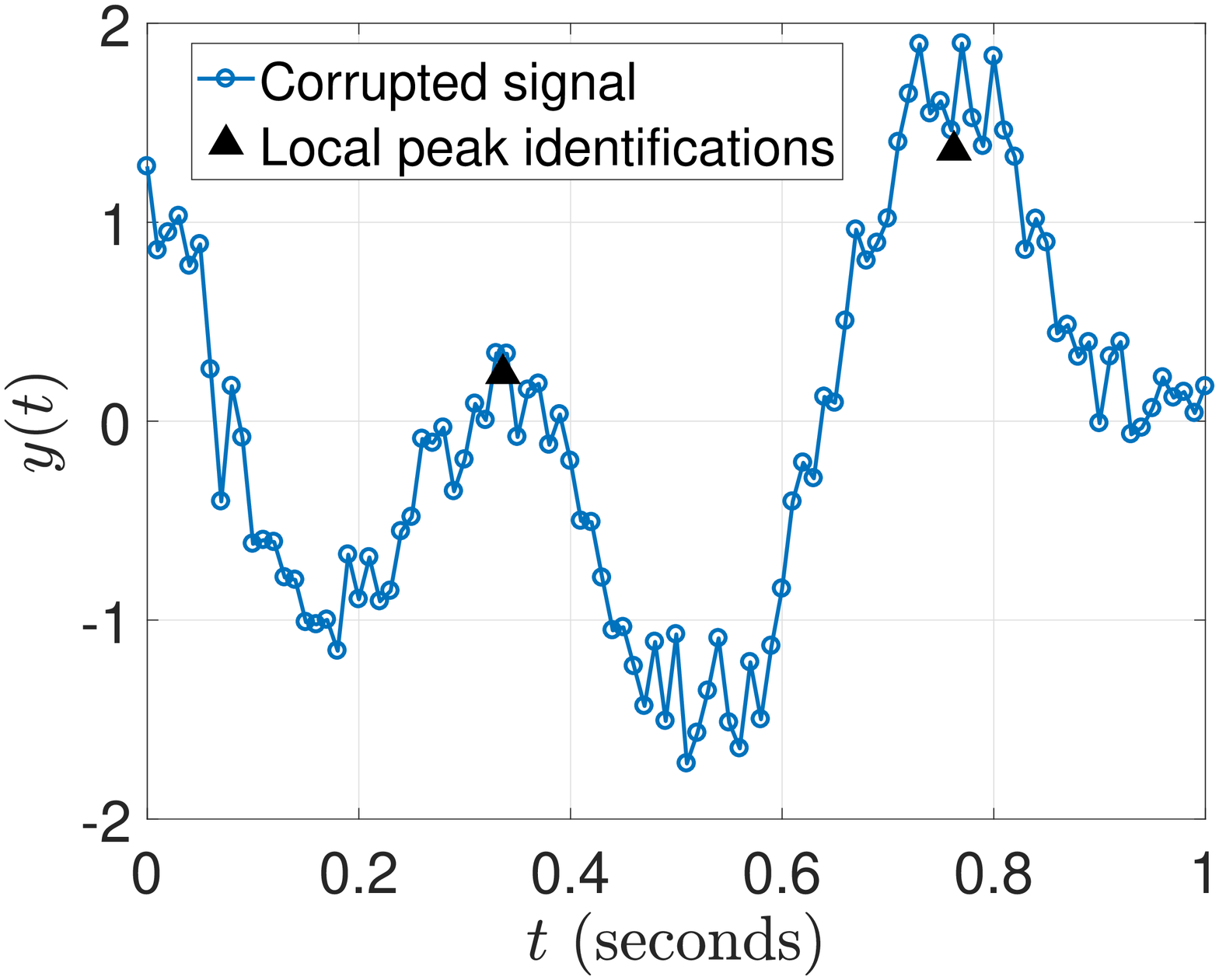}
        \caption{\label{fig:Peak_Detection_CorruptedSignal} Corrupted signal and its estimated local peaks.}
    \end{subfigure}
    \caption{An illustration of the peak identification algorithm.}
\end{figure}

We use a numerical example in the following to verify the effectiveness of the proposed local peak identification algorithm. Let $y(t) = \cos(2\pi f_1 t+0.2\pi)+\cos(2\pi f_2 t+0.3\pi)+n(t)$, where we set $f_1=1\,$Hz, $f_2=2.5\,$Hz, and $n(t)\sim\mathcal{N}(0,\sigma^2)$ is additive white Gaussian noise with zero mean and variance $\sigma^2$. The signal $y(t)$ is sampled at a rate of $100\,$Hz from time $t=0\,$s to $t=1\,$s. When the noise is absent, the true locations of the two local peaks are $t_1\approx 0.331\,$s and $t_2\approx 0.760\,$s and the estimates of our proposed local peak identification algorithm are $\hat{t}_1\approx 0.327\,$s and $\hat{t}_2\approx 0.763\,$s, as shown in Fig.~\ref{fig:Peak_Detection_OriginalSignal}. When the noise is present and $\sigma$ is set to $0.2$, the estimates are $\hat{t}_1\approx 0.336\,$s and $\hat{t}_2\approx 0.762\,$s, as shown in Fig.~\ref{fig:Peak_Detection_CorruptedSignal}. As we can see from the results, the estimated locations of the local peaks are very close to those of the actual peaks even when the signal is corrupted with the noise, which shows the effectiveness of the proposed local peak identification algorithm.

Then, the speed of the moving object can be estimated as $\hat{v} = \frac{0.54\lambda}{\hat{\tau}}$, where $0.54\lambda$ is the distance between the first local peak of $\Delta\rho^2_{E_{ix}}(d)$ and the origin, and $\hat{\tau}$ is the location of the first local peak of $\Delta\hat{\rho}_G(\tau)$. A median filter is then applied to the speed estimates to remove the outliers.

\subsection{Acceleration Estimator}

Acceleration can be calculated from $\hat{v}$ obtained in Section~\ref{subsec:movingspeed}. One intuitive method of acceleration estimation is to take the difference of two adjacent speed estimates and then divide the difference of the speeds by the difference of their measurement time. However, this scheme is not robust as it is likely to magnify the estimation noise. Instead, we leverage the fact that the acceleration values can be approximated as a piecewise linear function as long as there are enough speed estimates within a short duration. $\ell_1$ trend filter produces trend estimates that are smooth in the sense of being piecewise linear~\cite{Kim09l1} and is well suited to our purpose. Thus, we adopt an $\ell_{1}$ trend filter to extract the piecewise linear trend embedded in the speed estimation and then, estimate the accelerations by taking differential of the smoothed speed estimation.

Mathematically, let $\hat{v}[n]$ denote $\hat{v}(n\Delta T)$, where $\Delta T$ is the interval between two estimates, and let $\tilde{v}[n]$ denote the smoothed one. Then, $\tilde{v}[n]$ is obtained by solving the following unconstrained optimization problem:
\begin{equation}
\underset{\tilde{v}[n],\forall n}{\text{min}} \sum_{n=1}^N\! \left( \tilde{v}[n]\!-\!\hat{v}[n] \right)^2 \!+\! \lambda \!\sum_{n=2}^{N-1}\!\! \Big| \tilde{v}[n-1]\!-\!2\tilde{v}[n]\!+\!\tilde{v}[n+1] \Big|,
\end{equation}
where $\lambda\geq 0$ is the regularization parameter used to control the trade-off between
smoothness of $\tilde{v}[n]$ and the size of the residual $|\tilde{v}[n]-\hat{v}[n]|$. Then, we obtain the acceleration estimation as $\hat{a}[n] = \frac{\left(\tilde{v}[n]-\tilde{v}[n-1]\right)}{\Delta T}$. As shown in \cite{Kim09l1}, the complexity of the $\ell_{1}$ filter grows linearly with the length of the data and can be calculated in real-time on most platforms.

\subsection{Gait Cycle Estimator}
When the estimated speed is within a certain range, e.g., from $1\,m/s$ to $2\,m/s$, and the acceleration estimates are small, then WiSpeed starts to estimate the corresponding gait cycle. In fact, the process for walking a single step can be decomposed into three stages: lifting one leg off the ground, using the lifted leg to contact with the ground and pushing the body forward, and keeping still for a short period of time before the next step. The same procedure is repeated until the destination is reached.

In terms of speed, one cycle of walking consists of an acceleration stage followed by a deceleration stage. WiSpeed leverages the periodic pattern of speed changes for gait cycle estimation. More specifically, WiSpeed localizes the local peaks in the speed estimates corresponding to the moments with the largest speeds. To achieve peak localization, we use the persistence-based scheme presented in \cite{Kozlov15Persistence} to formulate multiple pairs of local maximum and local minimum, and the locations of the local maximum are considered as the peak locations. The time interval between every two adjacent peaks is computed as a gait cycle. Meanwhile, the moving distance between every two adjacent peaks is calculated as the estimation of the stride length.

\section{Experimental Results}
\label{sec: Exp evaluations}
In this section, we first introduce the indoor environment and system setups of the experiments. Then, the performance of WiSpeed is evaluated in two applications: human walking monitoring and human fall detection.
\subsection{Environment}
We conduct extensive experiments in a typical office environment, with floorplan shown in Fig.~\ref{fig:Experiment_Map}. The indoor space is occupied by desks, computers, shelves, chairs, and household appliances. The same WiFi devices as introduced in Section~\ref{sec: Theoretical Foundation of WiSpeed} are used during the experiments.

\subsection{Experimental Settings}
Two sets of experiments are performed. In the first set of experiments, we study the performance of WiSpeed in estimating the human walking speed. For device-free scenario, it shows that the number of steps and stride length can also be estimated besides the walking speed. Estimation accuracy is used as the metric which compares the estimated walking distances with the ground-truth distances, since measuring walking distance is much easier and accurate than measuring the speed directly. Different routes and locations of the devices are tested and the details of experiment setup are summarized in Tab.~\ref{tab: walking setting} and Tab.~\ref{tab:Moving_Antenna}. In the second set of experiments, we investigate the performance of WiSpeed as a human activity monitoring scheme. Two participants are asked to perform different activities, including standing up, sitting down, picking up things from the ground, walking, and falling down.
\begin{table}[!tbp]
\centering
\caption{Exp. settings for device-free human walking monitoring}
\label{tab: walking setting}
\begin{tabular}{|l|l|l|l|}
\hline
\diagbox{Setting}{Config.}    & Tx loc. & Rx loc. & Route index   \\ \hline
Setting \#1                   & Tx \#1      & Rx \#1      & Route \#1/\#2 \\ \hline
Setting \#2                   & Tx \#1      & Rx \#2      & Route \#1/\#2 \\ \hline
Setting \#3                   & Tx \#2      & Rx \#1      & Route \#1/\#2 \\ \hline
Setting \#4                   & Tx \#3      & Rx \#2      & Route \#3/\#4 \\ \hline
Setting \#5                   & Tx \#4      & Rx \#2      & Route \#3/\#4 \\ \hline
Setting \#6                   & Tx \#3      & Rx \#3      & Route \#3/\#4 \\ \hline
\end{tabular}
\end{table}

\begin{table}[!tbp]
\centering
\caption{Exp. settings for device-based speed monitoring}
\label{tab:Moving_Antenna}
\begin{tabular}{|l|l|l|l|}
\hline
\diagbox{Setting}{Config.}    & Tx loc. & Rx loc. & Route index   \\ \hline
Setting \#7                   & moving      & Rx \#1      & Route \#1/\#2 \\ \hline
Setting \#8                   & moving      & Rx \#4      & Route \#1/\#2 \\ \hline
Setting \#9                   & moving      & Rx \#1      & Route \#3/\#4 \\ \hline
Setting \#10                  & moving      & Rx \#4      & Route \#3/\#4 \\ \hline
\end{tabular}
\end{table}

\subsection{Human Walking Monitoring}

\begin{figure*}[tb]
    \centering
    \includegraphics[width=0.95\textwidth]{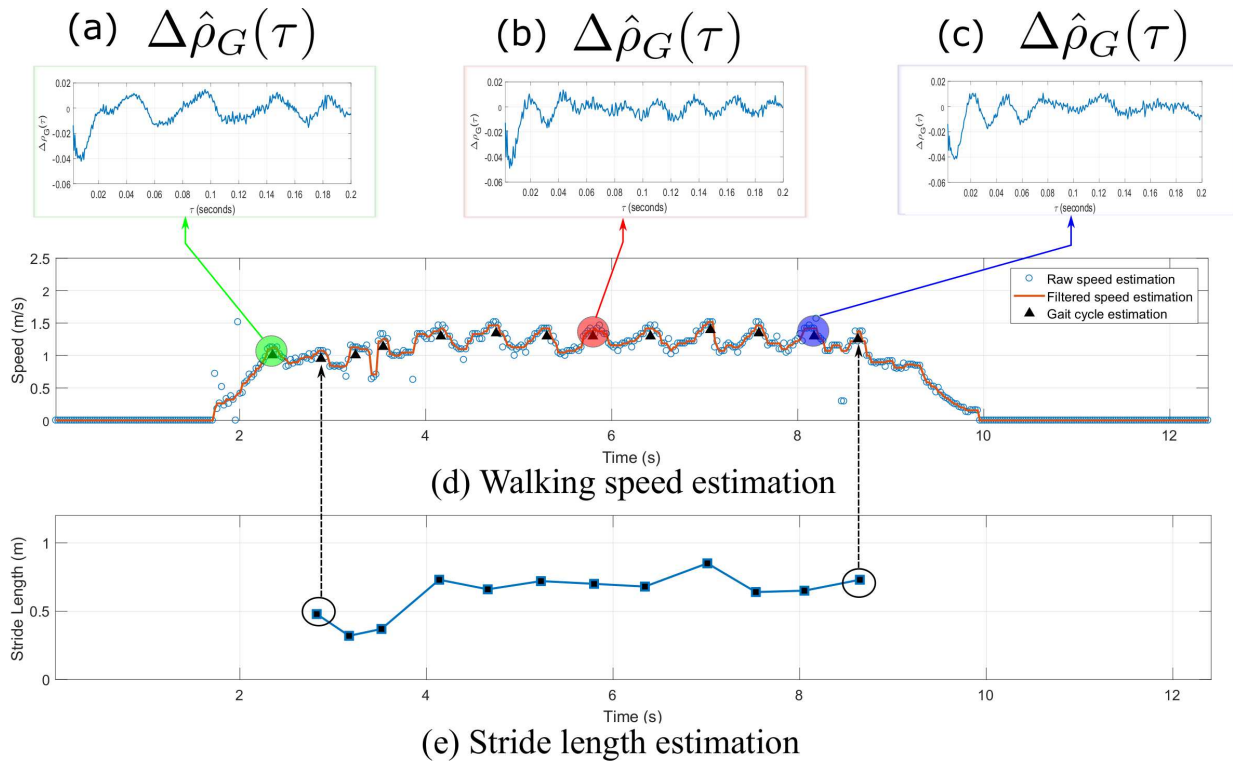}
    \caption{\label{fig:HumanWalkMonitor} Experimental results for human walking monitoring under Setting \#1 and Route \#1.}
\end{figure*}

Fig.~\ref{fig:HumanWalkMonitor} visualizes one of the experimental results under Setting \#1 of Route \#1, i.e., both the Tx and Rx are static and one experimenter walks along the specified route. Fig.~\ref{fig:HumanWalkMonitor}a--c show three snapshots of estimated ACFs at different time instances marked in Fig.~\ref{fig:HumanWalkMonitor}d. From Fig.~\ref{fig:HumanWalkMonitor}, we can conclude that although the ACFs are very different, the locations of the first local peak of $\Delta\hat{\rho}_G(\tau)$ are highly consistent as long as the ACFs are calculated under similar walking speeds.

Fig.~\ref{fig:HumanWalkMonitor}d shows the results of walking speed estimation for the experiment, and we can see a very clear pattern of walking due to the acceleration and deceleration. The corresponding stride length estimation is shown in Fig.~\ref{fig:HumanWalkMonitor}e. The estimated walking distance is $8.46\,$m and it is within $5.75\%$ of the ground-truth distance of $8\,$m. On the other hand, the average stride length is $0.7\,$m and very close to the average walking stride length of the participants.

\begin{figure}[tb]
    \centering
    \begin{subfigure}[b]{0.23\textwidth}
        \includegraphics[width=\textwidth]{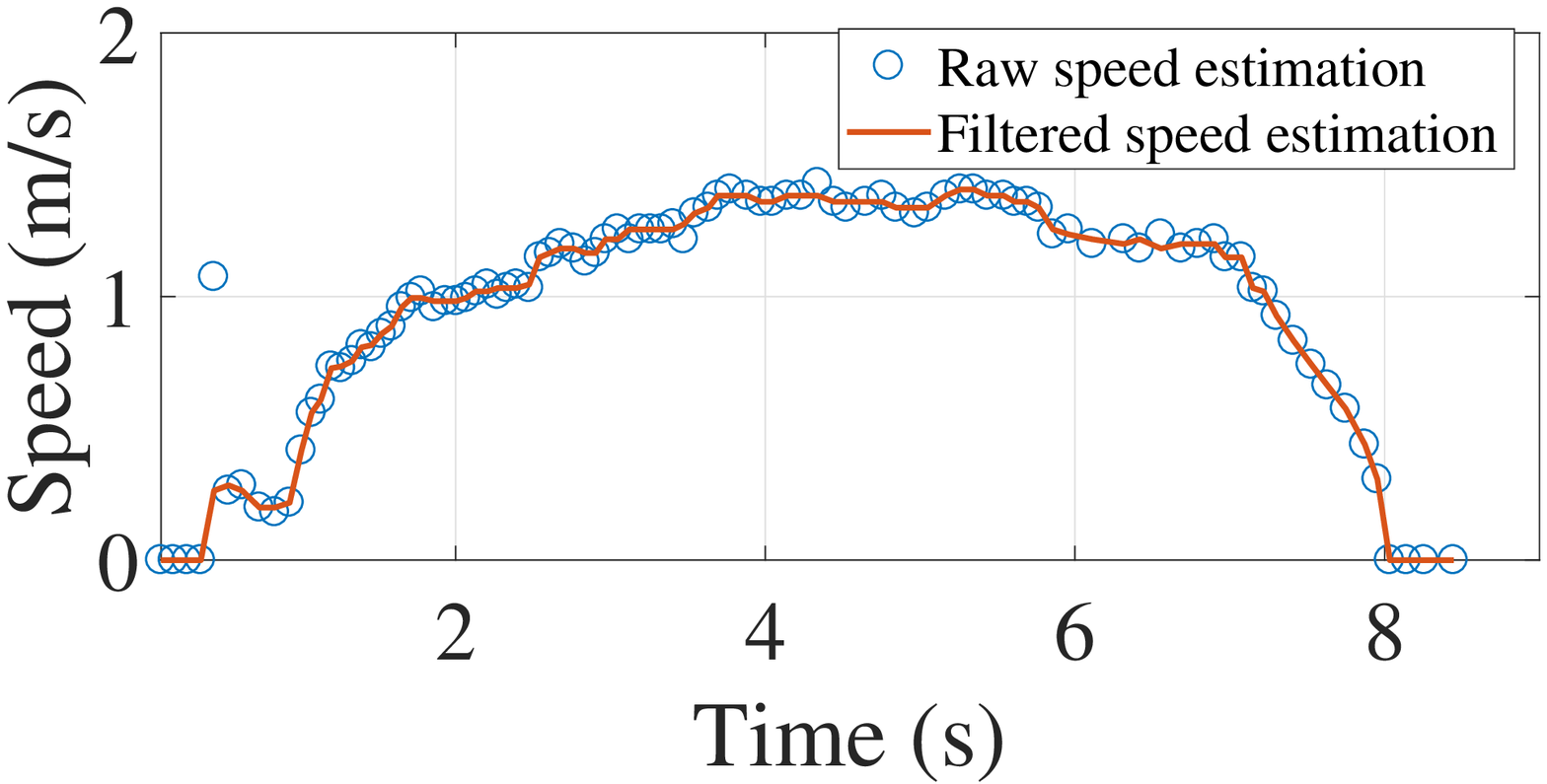}
        \caption{\label{fig:Active_Speed_Fast} Tx moves at a higher speed.}
    \end{subfigure}
    ~
    \begin{subfigure}[b]{0.23\textwidth}
        \includegraphics[width=\textwidth]{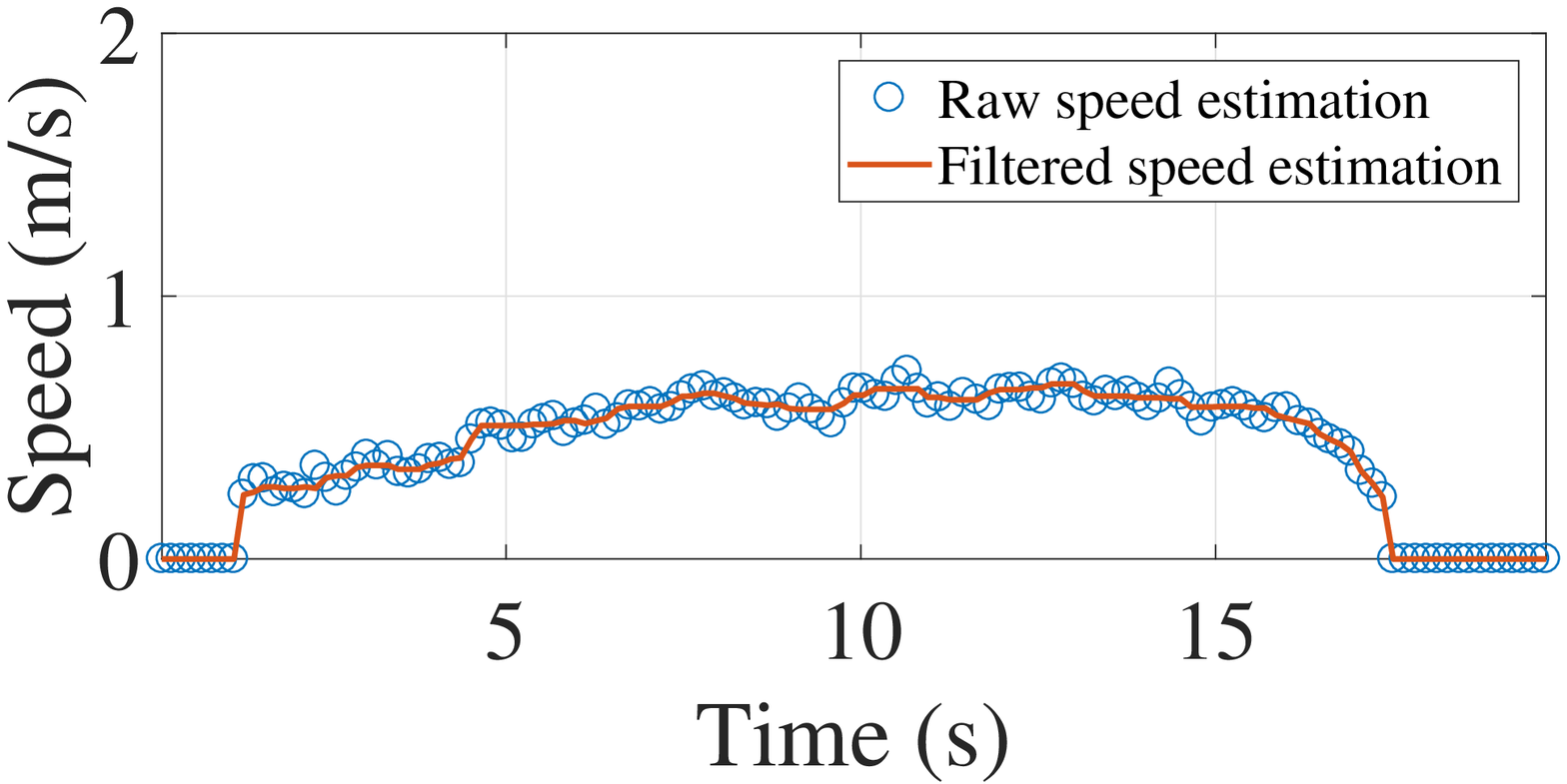}
        \caption{\label{fig:Active_Speed_Slow} Tx moves at a lower speed.}
    \end{subfigure}
    \caption{\label{fig:Active_Speed} Speed estimation for a moving Tx.}
\end{figure}

Fig.~\ref{fig:Active_Speed} shows two typical speed estimation results both under Setting \#7 of Route \#1 where the Tx is attached to a cart and one experimenter pushes the cart along the specified route. The cart moves at different speeds for these two realizations, and Fig.~\ref{fig:Active_Speed_Fast} and Fig.~\ref{fig:Active_Speed_Slow} show the corresponding speed estimates, respectively. As we can see from the estimated speed patterns, there are no periodic patterns like the device-free walking speed estimates as in Fig.~\ref{fig:HumanWalkMonitor}d. This is because when the Tx is moving, the energy of the EM waves reflected by the human body is dominated by that radiated by the transmit antennas and WiSpeed can only estimate the speed of moving antennas. The estimated moving distance for the case that Tx moves at a higher speed is $8.26\,$m and the other one is $8.16\,$m, where the ground-truth distance is $8\,$m.

\begin{figure*}[tb]
    \centering
    \begin{subfigure}[b]{0.23\textwidth}
        \includegraphics[width=\textwidth]{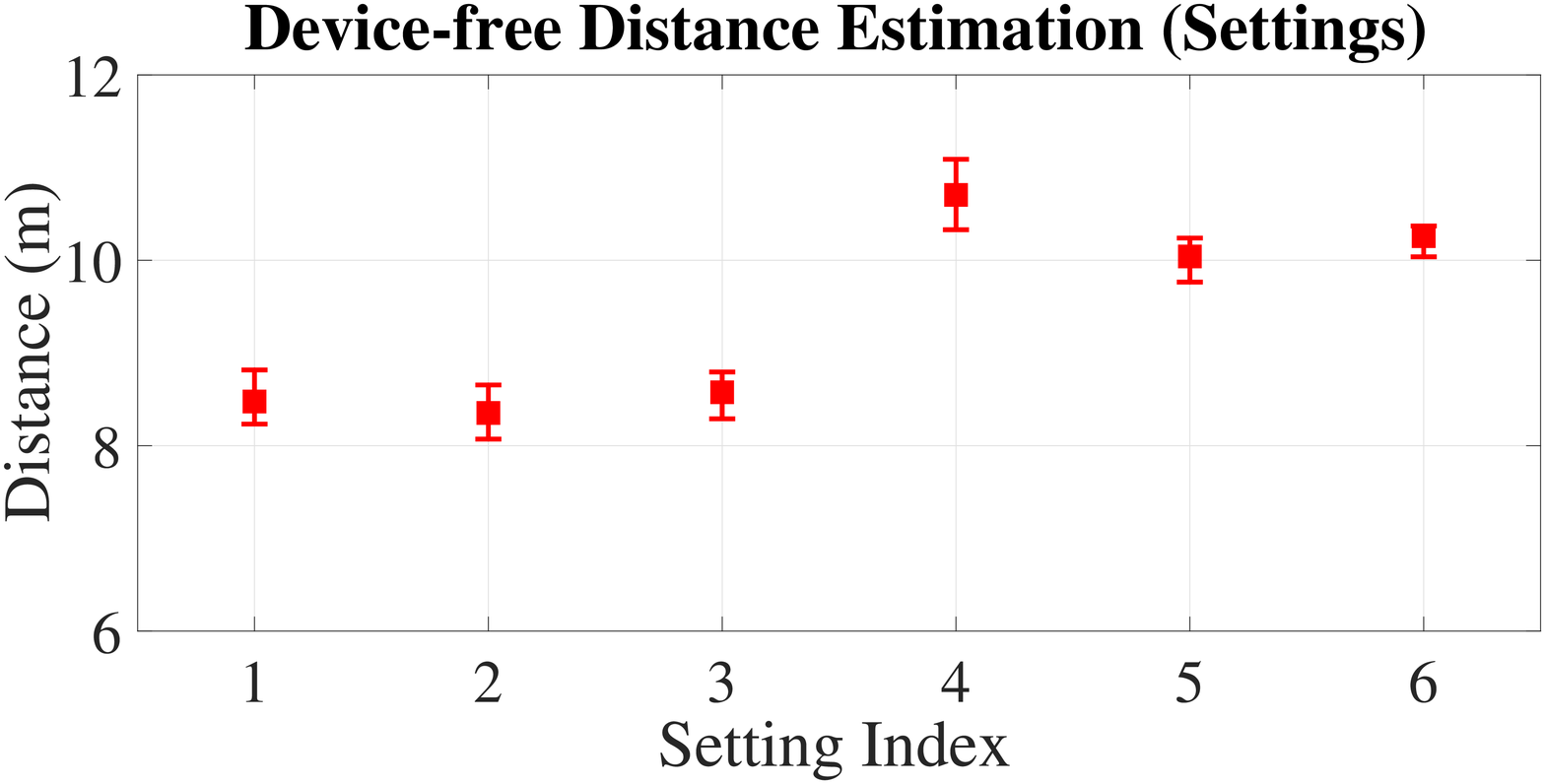}
        \caption{\label{fig:DF_Distance_Settings} Estimation results for Setting \#1 -- \#6.}
    \end{subfigure}
    ~
    \begin{subfigure}[b]{0.23\textwidth}
        \includegraphics[width=\textwidth]{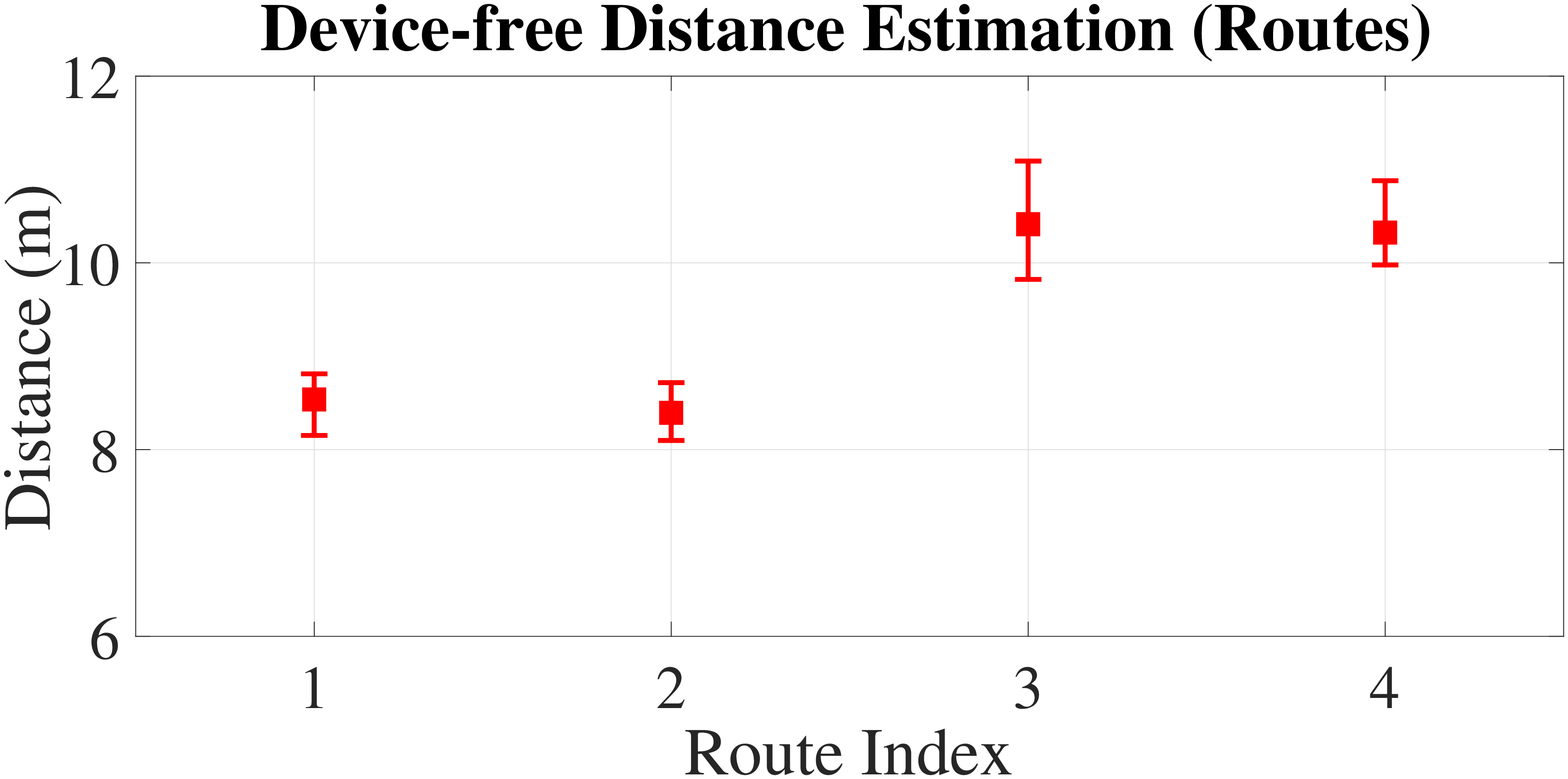}
        \caption{\label{fig:DF_Distance_Routes} Estimation results for Route \#1 -- \#4.}
    \end{subfigure}
    ~
    \begin{subfigure}[b]{0.23\textwidth}
        \includegraphics[width=\textwidth]{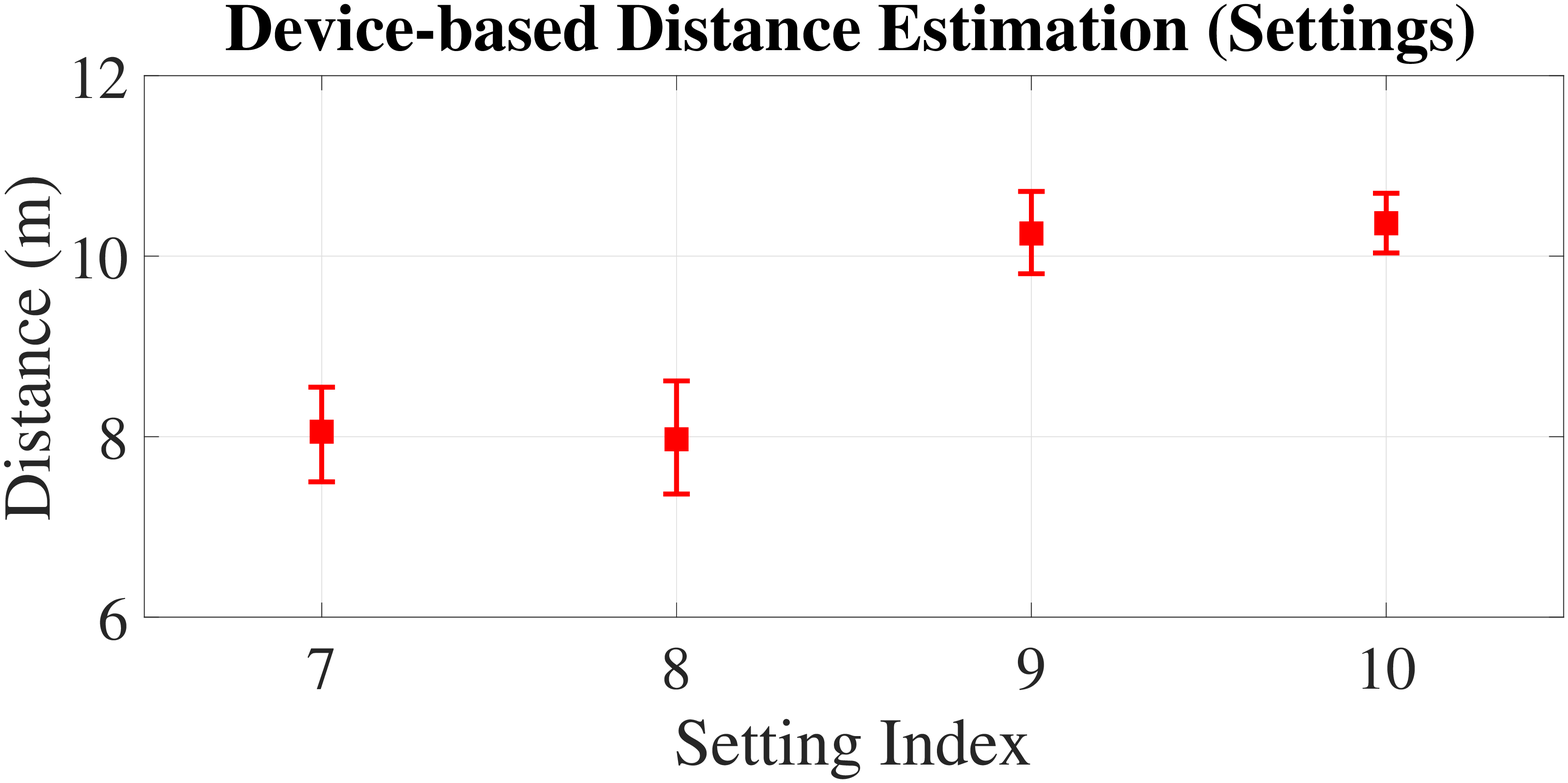}
        \caption{\label{fig:DB_Distance_Settings} Estimation results for Setting \#7 -- \#10.}
    \end{subfigure}
    ~
    \begin{subfigure}[b]{0.23\textwidth}
        \includegraphics[width=\textwidth]{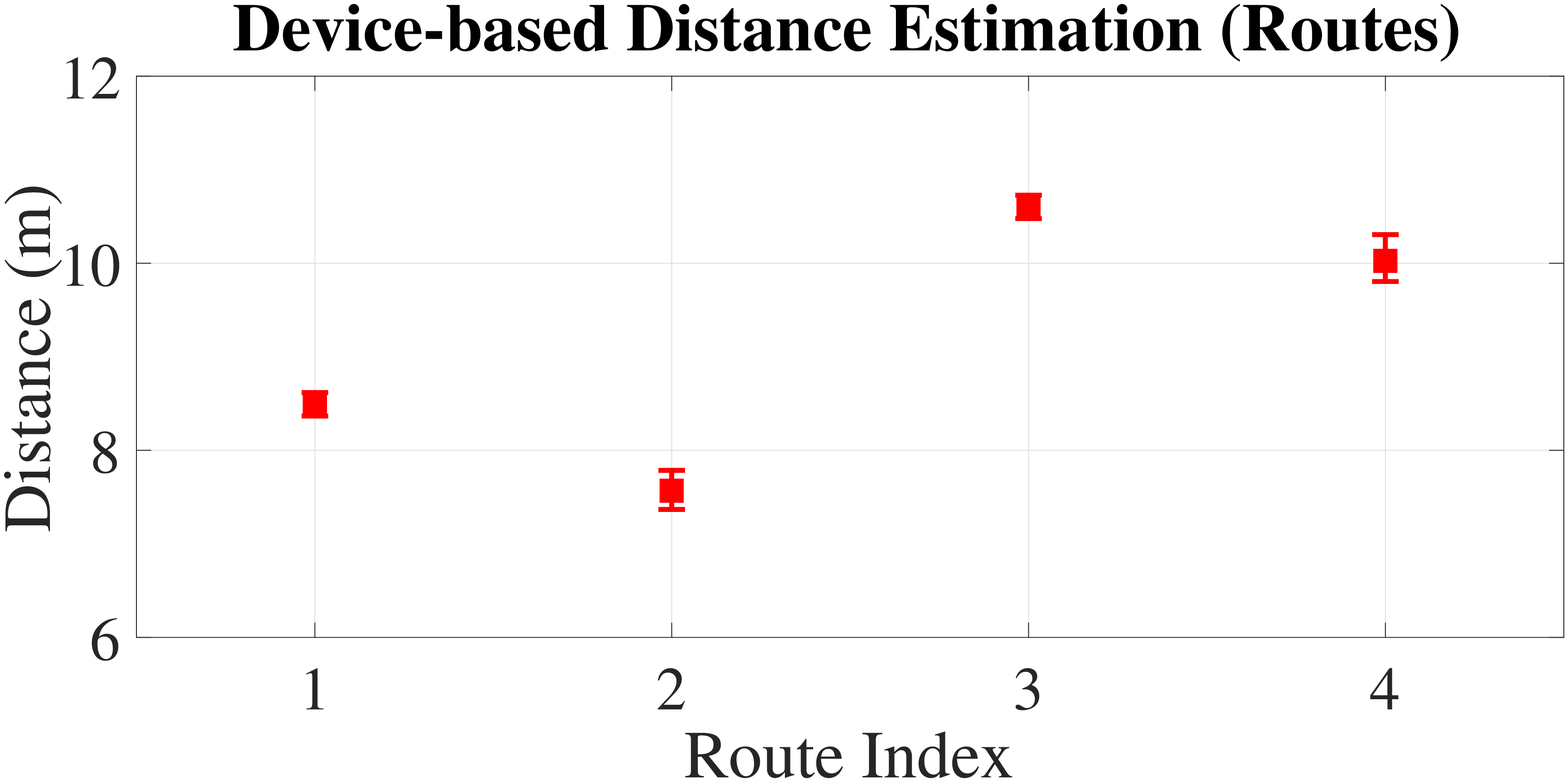}
        \caption{\label{fig:DB_Distance_Routes} Estimation results for Route \#1 -- \#4.}
    \end{subfigure}
    \caption{\label{fig:WalkingDistance} Error distribution of distance distance estimates under different conditions.}
\end{figure*}

Fig.~\ref{fig:WalkingDistance} summarizes the accuracy of the $200$ experiments of human walking speed estimation. More specifically, Fig.~\ref{fig:DF_Distance_Settings} shows the error distribution for Setting \#1 -- \#6, and Fig.~\ref{fig:DF_Distance_Routes} demonstrates the corresponding error distribution for Route \#1 -- \#4; Fig.~\ref{fig:DB_Distance_Settings} shows the error distribution for Setting \#7 -- \#10, and Fig.~\ref{fig:DB_Distance_Routes} demonstrates the corresponding error distribution for Route \#1 -- \#4. The bottom and top error bars stand for the $5\%$ percentiles and $95\%$ percentiles of the estimates, respectively, and the middle of point is the sample mean of the estimates. The ground-truths for Routes \#1--\#4 are shown in Fig.~\ref{fig:Experiment_Map}. From the results, we find that (i)  WiSpeed performs consistently for different Tx/Rx locations, routes, subjects, and walking speeds, indicating the robustness of WiSpeed under various scenarios, and (ii) WiSpeed tends to overestimate the moving distances under device-free settings. This is because  we use the route distances as baselines and ignore the displacement of the subjects in the direction of gravity. Since WiSpeed measures the absolute moving distance of the subject in the coverage area, the motion in the gravity direction would introduce a bias into the distance estimation.

In summary, WiSpeed achieves a MAPE of $4.85\%$ for device-free human walking speed estimation and $4.62\%$ for device-based speed estimation, which outperforms the existing approaches, even with only a single pair of WiFi devices and in severe NLOS conditions.

\subsection{Human Fall Detection}

\begin{figure*}[!tb]
    \centering
    \includegraphics[width=0.98\textwidth]{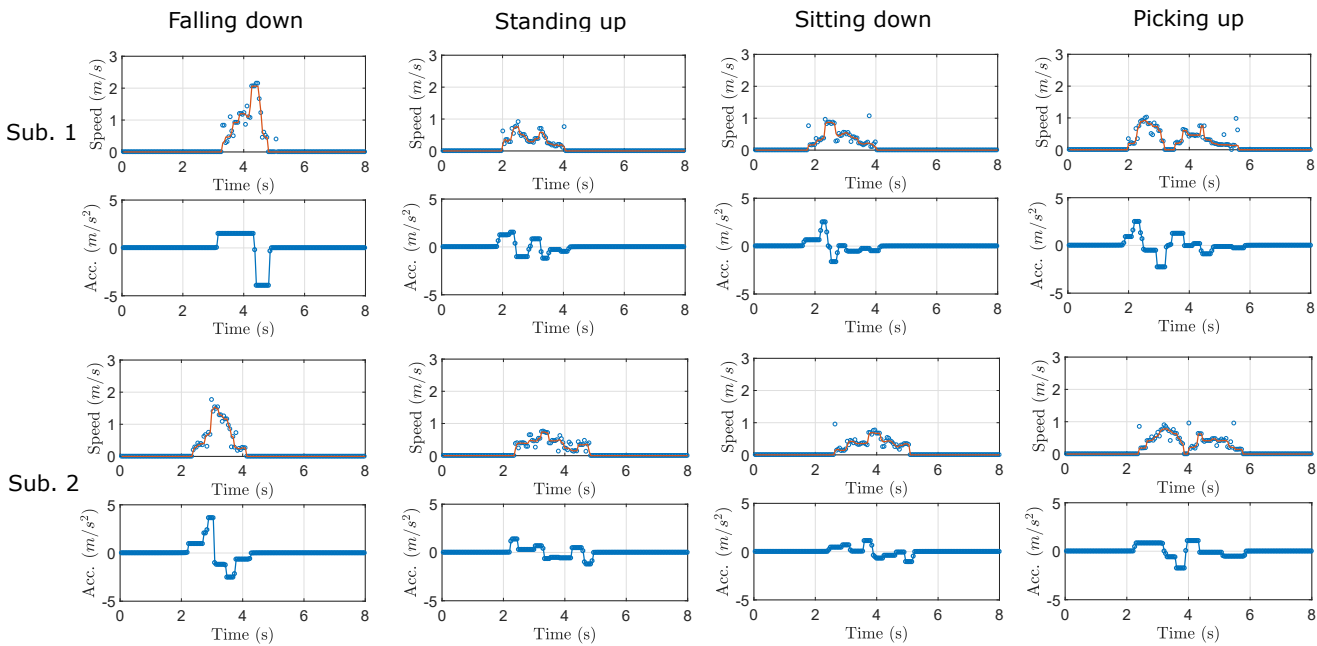}
    \caption{\label{fig:ActivityMonitor} Speed and Acceleration for different activities  and subjects.}
\end{figure*}

In this subsection, we show that WiSpeed can differentiate falling down from other normal daily activities. We collect a total of five sets of data: (i) falling to the ground, (ii) standing up from a chair, (iii) sitting down on a chair, and (iv) bowing and picking up items from the ground, (v) walking inside the room. Each experiment lasts for $8\,$s. We collect $20$ datasets of the falling down activity from two subjects, and $10$ datasets for each of the other four activities from the same two subjects. The experiments are conducted in Room \#$5$, and the WiFi Tx and Rx are placed at Location Tx \#$1$ and Rx \#$2$ as shown in Fig.~\ref{fig:Experiment_Map}. Fig.~\ref{fig:ActivityMonitor} shows a snapshot of speed and acceleration estimation results for different activities and subjects.

\begin{figure}[!tb]
    \centering
    \includegraphics[width=0.35\textwidth]{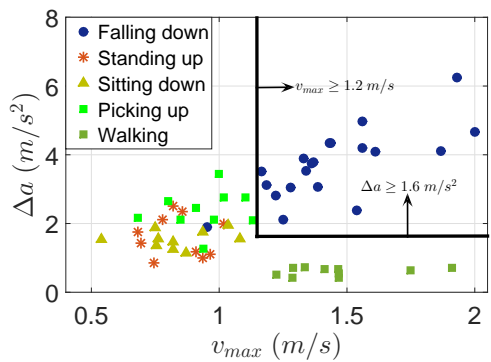}
    \caption{\label{fig:ActivityDetectionScatterplot} Distribution of the two metrics for all the activities.}
\end{figure}

Realizing that the duration of a real-world falling down can be as short as $0.5\,$s and the human body would experience a sudden acceleration and then a deceleration~\cite{Bagala12Evaluation}, we propose two metrics for falling down detection: (i) the maximum change in acceleration within $0.5\,$s, denoted as $\Delta a$, and (ii) the maximum speed during the period of the maximum change of acceleration, written as $v_{\mathrm{max}}$. Fig.~\ref{fig:ActivityDetectionScatterplot} shows the distribution of $(\Delta a, v_{\mathrm{max}})$ of all activities from the two subjects. Obviously, by setting two thresholds: $\Delta a\geq 1.6\,m/s^{2}$ and $v_{\max}\geq 1.2\,m/s$, WiSpeed could differentiate falls from the other four activities except one outlier, leading to a detection rate of $95\%$ and zero false alarm, while \cite{Wang14Eeyes} requires machine learning techniques. This is because WiSpeed extracts the most important physical features for activity classification, namely, the speed and the change of acceleration, while \cite{Wang14Eeyes} infers these two physical values indirectly.

\section{Discussion}
\label{sec:discussion}
In this section, we discuss the system parameter selections for different applications and their impact on the computational complexity of WiSpeed.

\subsection{Tracking a Fast Moving Object}
In order to track fast speed-varying object, we adopt the following equation with a reduced number of samples to calculate the sample auto-covariance function:
\begin{eqnarray}
\label{eq:newACF}
& & \!\!\!\!\!\!\!\!\!\!\hat{\gamma}_G(\tau,f) =  \nonumber \\
& & \!\!\!\!\!\!\!\!\!\!\!\!\!\!\frac{1}{M}\!\!\!\sum_{t=T-M+1}^{T}\!\!\!\!\!\! \left( G(t\!-\!\tau,\!f)\!-\!\bar{G}(f) \right)\!\!\left( G(t,\!f)\!-\!\bar{G}(f) \right),
\end{eqnarray}
where $T$ is the length of the window, $M$ is the number of samples for averaging, and $\bar{G}(f)$ is the sample average. \eqref{eq:newACF} shows that to estimate a moving subject with speed $v$, WiSpeed requires a time window with a duration $T_0=\frac{0.54\lambda}{v}+\frac{M}{F_s}$ seconds. Essentially, WiSpeed captures the average speed of motion in a period of time rather than the instantaneous moving speed. For instance, with $v=1.3\,m/s$, $F_s=1500\,$Hz, $f_c= 5.805\,$GHz, and $M=100$, $T_{0}$ is around $0.12\,$s. In case that the speed changes significantly within a duration of $T_{0}$, the performance of WiSpeed would degrade. To track the speed of a fast-varying moving subject, a smaller $T_{0}$ is desirable, which can be achieved by increasing the channel sampling rate $F_{s}$ or increasing the carrier frequency to reduce the wavelength $\lambda$.

\subsection{Computational Complexity} The main computational complexity comes from the estimation of the overall ACF $\hat{\rho}_G(\tau)$, giving rise to a total of $FMT_0F_s$  multiplications where $F$ is the number of available subcarriers. For motions with slow-varying speeds such as walking and standing up, a lower channel sampling rate suffices which could reduce the complexity. For example, in our experiments of human walking speed estimation and human fall detection, $F_s=1500\,$Hz, $f_c=5.805\,$GHz, $F=180$, and $M=100$, the total number of multiplications for WiSpeed to produce one output is around $3$ million. This leads to a computational time of $80.4\,$ms on a desktop with Intel Core i7-7500U processor and $16$GB memory, which is short enough for real-time applications.

\section{Conclusions}
\label{sec: Conclusions}
In this work, we propose WiSpeed, a universal indoor speed estimation system for human motions leveraging commercial WiFi, which can estimate the speed of a moving object under either device-free or device-based condition. WiSpeed is built upon the statistical theory of EM waves which quantifies the impact of human motions on EM waves for indoor environments. We conduct extensive experiments in a typical indoor environment which demonstrates that WiSpeed can achieve a MAPE of $4.85\%$ for device-free human walking speed monitoring and a MAPE of $4.62\%$ for device-based speed estimation. Meanwhile, it achieves an average detection rate of $95\%$ with no false alarms for human fall detection. Due to its large coverage, robustness, low cost, and low computational complexity, WiSpeed is a very promising candidate for indoor passive human activity monitoring systems.

\section*{Appendix}
\subsection{Derivation of \eqref{eq: cov of G}}
\label{App:1}
First, we can rewrite $G(t,f)$ as
\begin{eqnarray}
G(t,f) = \sum_{u\in\{x,y,z\}} G_u(t,f) +\varepsilon(t,f),
\end{eqnarray}
where $G_u(t,f)\triangleq |E_{su}(f)|^2 + 2\mathrm{Re}\left\{E^*_{su}(f)\sum_{i\in\Omega_d}E_{iu}(t,f)\right\}\nonumber+\left|\sum_{i\in\Omega_d}E_{iu}(t,f)\right|^2$. Then, the covariance of $G(t,f)$ can be written as
\begin{eqnarray}
& &\gamma_G(\tau,f) = \mathrm{cov} \big(G(t,f),G(t-\tau,f)\big)  \nonumber\\
&=& \!\!\!\!\!\!\!\!\sum_{u\in\{x,y,z\}}\!\!\!\!\! \mathrm{cov} \!\big(\! G_u(t\!,\!f)\!,\!G_u(t\!-\!\tau\!,\!f) \!\big)\! \!+\!\! \mathrm{cov}\!\big(\! \varepsilon(t\!,\!f)\!,\!\varepsilon(t\!-\!\tau\!,\!f) \!\big)\!  \nonumber \\
&=& \!\!\!\!\!\!\!\!\sum_{u\in\{x,y,z\}}\!\!\!\!\!\! \mathrm{cov} \big( G_u(t,f),G_u(t\!-\!\tau,f) \big) \!\!+\!\! \delta(\tau)\sigma^2(f),
\end{eqnarray}
which is due to Assumptions~\ref{Assump: 2}-\ref{Assump: 3} and the assumptions of the noise term. Thus, in the following, we only need to focus on the term $\gamma_{G_u}(\tau,f)\triangleq\mathrm{cov}\big( G_u(t,f),G_u(t-\tau,f) \big)$, that is, for $\forall u\in\{x,y,z\}$, we have the equation \eqref{eq:decomposition_gama}.
\begin{figure*}
\begin{eqnarray}
\label{eq:decomposition_gama}
& & \gamma_{G_u}(\tau,f) = \Big\langle G_u(t,f)-\langle G_u(t,f) \rangle, G_u(t-\tau,f)-\langle G_u(t-\tau,f) \rangle \Big\rangle \nonumber \\
&=& \bigg\langle \underbrace{2\mathrm{Re}\big\{ E^*_{su}(f)\sum_{i\in\Omega_d}E_{iu}(t,f) \big\}}_{\mathcal{A}_1}+\underbrace{\Big( \big|\sum_{i\in\Omega_d}E_{iu}(t,f)\big|^2 -\langle \big| \sum_{i\in\Omega_d}E_{iu}(t,f) \big|^2 \rangle \Big)}_{\mathcal{A}_2},  \nonumber\\
& & \!\!\!\!\!\!\!\! \underbrace{2\mathrm{Re}\big\{ E^*_{su}(f)\sum_{i\in\Omega_d}E_{iu}(t-\tau,f) \big\}}_{\mathcal{A}_3}+\underbrace{\Big( \big|\sum_{i\in\Omega_d}E_{iu}(t-\tau,f)\big|^2 -\langle \big| \sum_{i\in\Omega_d}E_{iu}(t-\tau,f) \big|^2 \rangle \Big)}_{\mathcal{A}_4}\bigg\rangle.
\end{eqnarray}
\rule{\textwidth}{0.5pt}
\end{figure*}
We begin with the term $\Big\langle \mathcal{A}_1,\mathcal{A}_3 \Big\rangle$. For notational convenience, define $E_{iu}(t,f)\triangleq a_i(t)+j b_i(t)$ and $E_{su}(f) \triangleq u+j v$, for $\forall i\in\Omega_d$, $\forall u\in\{x,y,z\}$, and $a_i$, $b_i$, $u$, $v$ are all real. Then, we have
\begin{eqnarray}
& &\Big\langle \mathcal{A}_1,\mathcal{A}_3 \Big\rangle \nonumber\\
\!\!\!\!\!\!\!\!&=&\!\!\!\!\!  4\Big\langle\!\! u\!\!\sum_{i\in\Omega_d}\! a_i(t)\!\!+\!\!v\!\!\sum_{i\in\Omega_d}\! b_i(t),\! u\!\!\sum_{i\in\Omega_d}\!\!a_i(t\!\!-\!\!\tau\!)\!\!+\!\!v\!\!\sum_{i\in\Omega_d}\!\!b_i(t\!\!-\!\!\tau)\!\! \Big\rangle\nonumber \\
\!\!\!\!\!\!\!\!&=&\!\!\!\!\! 4 u^2\!\! \sum_{i\in\Omega_d}\!\! \Big\langle\!\! a_i(t),\!a_i(t\!-\!\tau)\!\! \Big\rangle \!\!+\!\! 4v^2\!\!\sum_{i\in\Omega_d}\!\!\Big\langle\! b_i(t),b_i(t-\tau)\! \Big\rangle \nonumber \\
\!\!\!\!\!\!\!\!&=&\!\!\!\!\! 4(u^2+v^2)\sum_{i\in\Omega_d} \Big\langle a_i(t),a_i(t-\tau) \Big\rangle,
\end{eqnarray}
where we apply the assumption that the real and imaginary parts of the electric field have the same statistical behaviors. At the same time, we have
\begin{eqnarray}
& &\mathrm{cov}(E_{iu}(t,f),E_{iu}(t-\tau,f)) \nonumber\\
&=&  \Big\langle E_{iu}(t,f),E_{iu}(t-\tau,f) \Big\rangle\nonumber \\
&=& \Big\langle a_i(t),a_i(t-\tau) \Big\rangle + \Big\langle b_i(t),b_i(t-\tau) \Big\rangle \nonumber\\
&=& 2\Big\langle a_i(t),a_i(t-\tau) \Big\rangle.
\end{eqnarray}
Thus, we have
\begin{eqnarray}
& & \Big\langle \mathcal{A}_1,\mathcal{A}_3 \Big\rangle\nonumber\\
\!\!\!\!\!\!\!\!\!\!&=&\!\!\!\!\! 2|E_{su}(f)|^2\!\!\sum_{i\in\Omega_d}\!\! \mathrm{cov}\big( E_{iu}(t,\!f)\!,\! E_{iu}(t\!-\!\tau,\!f)\! \big).
\end{eqnarray}
Next, we derive the term $\Big\langle \mathcal{A}_1,\mathcal{A}_4 \Big\rangle$ as shown in \eqref{eq:term1_term4}.
\begin{figure*}
\begin{eqnarray}
\label{eq:term1_term4}
\Big\langle \mathcal{A}_1,\mathcal{A}_4 \Big\rangle &=& 2\Big\langle u\sum_{i\in\Omega_d}a_i(t)\!+\!v\sum_{i\in\Omega_d}b_i(t), \Big(\sum_{i\in\Omega_d} a_i(t\!-\!\tau)\Big)^2 \!+\! \Big(\sum_{i\in\Omega_d} b_i(t\!-\!\tau)\Big)^2 \!-\!\langle \big| \sum_{i\in\Omega_d}E_{iu}(t\!-\!\tau,f) \big|^2 \rangle  \Big\rangle \nonumber \\
&=& 2\Big\langle u\sum_{i\in\Omega_d}a_i(t)+v\sum_{i\in\Omega_d}b_i(t), \Big(\sum_{i\in\Omega_d} a_i(t-\tau)\Big)^2+ \Big(\sum_{i\in\Omega_d} b_i(t-\tau)\Big)^2 \Big\rangle \nonumber \\
&=& 2u\sum_{i\in\Omega_d} \Big\langle a_i(t),a_i^2(t-\tau) \Big\rangle +2v\sum_{i\in\Omega_d}\Big\langle b_i(t),b^2_i(t-\tau) \Big\rangle.
\end{eqnarray}
\end{figure*}
According to the integral representation of the electric field in \eqref{eq: Integral representation}, we have
\begin{eqnarray}
\!\!\!\!& &|E_{iu}(t,f)|^2 \qquad\qquad\qquad\qquad\qquad\qquad\qquad\qquad\!\!\!\!\nonumber\\
\!\!\!\!\!\!\!\!\!\!\!\!\!\!&=&\!\!\!\!\!\!\!\! \iint_{4\pi}\!\!\!\!\!\!  F_{iu}\!(\!\Theta_1\!) \!F^*_{iu}\!(\!\Theta_2\!) \!\exp(\!-\!j(\!\vec{k}(\!\Theta_1\!)\!\!-\!\!\vec{k}(\!\Theta_2\!)\!)\!\!\cdot\!\!\vec{v}_i t) \mathrm{d}\Theta_1\!\mathrm{d}\Theta_2,
\end{eqnarray}
and thus, the covariance between $E_{iu}(t,f)$ and $|E_{iu}(t-\tau,f)|^2$ can be expressed as
\begin{eqnarray}
\!\!\!\!\!\!\!& & \mathrm{cov}(E_{iu}(t,f),|E_{iu}(t-\tau,f)|^2)  \nonumber \\
\!\!\!\!\!\!\!&=&\!\!\!\!\Big\langle\!\! E_{iu}(t,f)\!\!-\!\!\langle E_{iu}(t,\!f) \rangle\!,\! |E_{iu}(t\!\!-\!\!\tau,f)|^2\!\!-\!\!\langle |E_{iu}(t\!\!-\!\!\tau,\!f)|^2\! \rangle\!\! \Big\rangle \nonumber \\
&=&\Big\langle E_{iu}(t,f), |E_{iu}(t-\tau,f)|^2 \Big\rangle \nonumber \\
&=&\!\!\!\!\iiint_{4\pi}\!\!\!\! \Big\langle \!F_{iu}(\Theta_1)\!,\!F_{iu}(\!\Theta_{21}\!)F^*_{iu}(\!\Theta_{22}\!) \Big\rangle \exp(-j\vec{k}(\Theta_1)\!\cdot\!\vec{v}_i t) \nonumber \\
& & \quad \exp(\!-\!j(\vec{k}(\!\Theta_{21}\!)\!\!-\!\!\vec{k}(\!\Theta_{22})\!)\!\cdot\!\vec{v}_i (t\!-\!\tau))\,\mathrm{d}\Theta_1\,\mathrm{d}\Theta_{21}\,\mathrm{d}\Theta_{22} \nonumber \\
&=&\!\!\!\! \int_{4\pi} \Big\langle F_{iu}(\Theta_1),|F_{iu}(\Theta_1)|^2 \Big\rangle \exp(-j\vec{k}(\Theta_1)\cdot\vec{v}_i t)\,\mathrm{d}\Theta_1 \nonumber \\
&=&\!\!\!\! \int_{4\pi} \bigg(\Big\langle \mathrm{Re}\Big\{F_{iu}(\Theta_1)\Big\},\mathrm{Re}\Big\{F_{iu}(\Theta_1)\Big\}^2 \Big\rangle \!\!+\!\!\nonumber\\
& &\!\!\!\!\!\! j\Big\langle \!\mathrm{Im}\Big\{\!F_{iu}(\!\Theta_1\!)\!\Big\}\!,\!\mathrm{Im}\Big\{\!F_{iu}(\Theta_1)\!\Big\}^2 \!\Big\rangle\!\!\!\bigg)\!\! \exp(\!-\!j\vec{k}(\Theta_1)\!\cdot\!\vec{v}_it)\mathrm{d}\Theta_1 \nonumber \\
&=& 0,
\end{eqnarray}
since $\langle X^3 \rangle\equiv 0$ for any Gaussian random variable with zero mean. At the same time, we have
\begin{eqnarray}
& & \Big\langle E_{iu}(t,f), |E_{iu}(t-\tau,f)|^2 \Big\rangle \nonumber\\
&=& \!\!\!\!\Big\langle a_i(t),a^2_i(t\!-\!\tau) \Big\rangle \!+\! j\Big\langle b_i(t),b_i^2(t\!-\!\tau) \Big\rangle,
\end{eqnarray}
and thus, we have $\Big\langle a_i(t),a^2_i(t-\tau) \Big\rangle=0$. Plugging this result in \eqref{eq:term1_term4}, we can obtain
\begin{eqnarray}
\Big\langle \mathcal{A}_1,\mathcal{A}_4 \Big\rangle = 0.
\end{eqnarray}
Similarly, we can also derive that $\Big\langle \mathcal{A}_2,\mathcal{A}_3 \Big\rangle = 0$. At last, we derive the term $\Big\langle \mathcal{A}_2,\mathcal{A}_4 \Big\rangle$ as shown in \eqref{eq:term2_term_4}.
\begin{figure*}
\begin{eqnarray}
\label{eq:term2_term_4}
\Big\langle \mathcal{A}_2,\mathcal{A}_4 \Big\rangle &=& \mathrm{cov}\bigg( \Big( \sum_{i\in\Omega_d}a_i(t) \Big)^2+\Big( \sum_{i\in\Omega_d}b_i(t) \Big)^2, \Big( \sum_{i\in\Omega_d}a_i(t-\tau) \Big)^2+\Big( \sum_{i\in\Omega_d}b_i(t-\tau) \Big)^2 \bigg) \nonumber \\
&=& \mathrm{cov}\bigg( \Big( \sum_{i\in\Omega_d}a_i(t) \Big)^2,\Big( \sum_{i\in\Omega_d}a_i(t-\tau) \Big)^2 \bigg) + \mathrm{cov}\bigg( \Big( \sum_{i\in\Omega_d}b_i(t) \Big)^2,\Big( \sum_{i\in\Omega_d}b_i(t-\tau) \Big)^2 \bigg) \nonumber \\
&=& 2\sum_{i_1,i_2\in\Omega_d} \mathrm{cov}\bigg( a_{i1}(t) a_{i2}(t), a_{i1}(t-\tau)a_{i2}(t-\tau) \bigg) \nonumber \\
&=& 2\sum_{i\in\Omega_d} \mathrm{cov}\bigg( a^2_i(t), a^2_i(t-\tau)\bigg) +2\sum\limits_{\substack{i_1,i_2\in\Omega_d\\i_1\neq i_2}} \mathrm{cov}\bigg( a_{i1}(t) a_{i2}(t), a_{i1}(t-\tau)a_{i2}(t-\tau) \bigg).
\end{eqnarray}
\rule{\textwidth}{0.5pt}
\end{figure*}
Since for any two Gaussian random variables, $X$ and $Y$, with zero mean, the expectations can be evaluated by using of the following relationship~\cite{Papoulis02Probability}:
\begin{eqnarray}
\big\langle X^2 Y^2 \big\rangle = \big\langle X^2 \big\rangle \big\langle Y^2 \big\rangle + 2\big\langle XY \big\rangle^2,
\end{eqnarray}
then, we have, $\forall i\in\Omega_d$,
\begin{eqnarray}
& &\mathrm{cov}\big( a^2_i(t),a^2_i(t-\tau) \big) \nonumber\\
&=& \Big\langle a^2_i(t)-\big\langle a^2_i(t) \big\rangle, a_i^2(t-\tau)-\big\langle a_i^2(t-\tau) \big\rangle\Big\rangle \nonumber \\
&=& \big\langle a^2_i(t),a^2_i(t-\tau) \big\rangle - \big\langle a^2_i(t) \big\rangle \big\langle a^2_i(t-\tau) \big\rangle  \nonumber \\
&=& 2\big\langle a_i(t),a_i(t-\tau) \big\rangle^2 \nonumber \\
&=& \frac{1}{2} \mathrm{cov}\big( E_{iu}(t,f), E_{iu}(t-\tau,f) \big)^2.
\end{eqnarray}
For $i_1,i_2\in\Omega_d$ and $i_1\neq i_2$, we have
\begin{eqnarray}
& & \mathrm{cov}\big( a_{i1}(t)a_{i2}(t),a_{i1}(t-\tau)a_{i2}(t-\tau) \big) \nonumber\\
&=& \Big\langle a_{i1}(t)a_{i2}(t),a_{i1}(t-\tau)a_{i2}(t-\tau) \Big\rangle  \nonumber \\
&=& \Big\langle a_{i1}(t)a_{i1}(t-\tau), a_{i2}(t)a_{i2}(t-\tau) \Big\rangle \nonumber\\
&=& \Big\langle a_{i1}(t), a_{i1}(t-\tau)\Big\rangle \Big\langle a_{i2}(t), a_{i2}(t-\tau) \Big\rangle \nonumber\\
&=& \frac{1}{4} \mathrm{cov}\big( E_{i_1u}(t,f),E_{i_1u}(t-\tau,f) \big)\nonumber\\
& & \mathrm{cov}\big( E_{i_2u}(t,f),E_{i_2u}(t-\tau,f) \big).
\end{eqnarray}
Therefore, $\Big\langle \mathcal{A}_2,\mathcal{A}_4 \Big\rangle$ can be derived as
\begin{eqnarray}
\Big\langle \mathcal{A}_2,\mathcal{A}_4 \Big\rangle \!\!\!&=&\!\!\! \sum\limits_{\substack{i_1,i_2\in\Omega_d\\i_1\geq i_2}} \mathrm{cov}\big( E_{i_1u}(t,f),E_{i_1u}(t-\tau,f) \big)\nonumber\\
& &\mathrm{cov}\big( E_{i_2u}(t,f),E_{i_2u}(t-\tau,f) \big).
\end{eqnarray}
Finally, we can obtain the result shown in \eqref{eq: cov of G}.

\bibliographystyle{ieeetr}
\bibliography{main}

\begin{thebibliography}{10}

\bibitem{khan16IoT}
M.~Khan, B.~N. Silva, and K.~Han, ``Internet of things based energy aware smart
  home control system,'' {\em IEEE Access}, vol.~4, pp.~7556--7566, 2016.

\bibitem{Pinto17Wecare}
S.~Pinto, J.~Cabral, and T.~Gomes, ``We-care: An {IoT}-based health care system
  for elderly people,'' in {\em IEEE International Conference on Industrial
  Technology (ICIT)}, pp.~1378--1383, March 2017.

\bibitem{Schaefer16Fitness}
S.~E. Schaefer, C.~C. Ching, H.~Breen, and J.~B. German, ``Wearing, thinking,
  and moving: testing the feasibility of fitness tracking with urban youth,''
  {\em American Journal of Health Education}, vol.~47, no.~1, pp.~8--16, 2016.

\bibitem{Wang10Machine}
L.~Wang, G.~Zhao, L.~Cheng, and M.~Pietik{\"a}inen, {\em Machine learning for
  vision-based motion analysis: Theory and techniques}.
\newblock Springer, 2010.

\bibitem{Gurbuz16Micro}
S.~Z. Gurbuz, C.~Clemente, A.~Balleri, and J.~J. Soraghan,
  ``Micro-{Doppler}-based in-home aided and unaided walking recognition with
  multiple radar and sonar systems,'' {\em IET Radar, Sonar Navigation},
  vol.~11, no.~1, pp.~107--115, 2017.

\bibitem{Hsu17Gait}
C.-Y. Hsu, Y.~Liu, Z.~Kabelac, R.~Hristov, D.~Katabi, and C.~Liu, ``Extracting
  gait velocity and stride length from surrounding radio signals,'' in {\em
  Proc. of CHI Conference on Human Factors in Computing Systems},
  pp.~2116--2126, ACM, 2017.

\bibitem{Qian17WiDar}
K.~Qian, C.~Wu, Z.~Yang, Y.~Liu, and K.~Jamieson, ``Widar: Decimeter-level
  passive tracking via velocity monitoring with commodity wi-fi,'' in {\em
  Proc. of the 18th ACM International Symposium on Mobile Ad Hoc Networking and
  Computing}, p.~6, ACM, 2017.

\bibitem{Halperin11tool}
D.~Halperin, W.~Hu, A.~Sheth, and D.~Wetherall, ``Tool release: Gathering
  802.11n traces with channel state information,'' {\em SIGCOMM Comput. Commun.
  Rev.}, vol.~41, pp.~53--53, Jan. 2011.

\bibitem{Abdelnasser15WiGest}
H.~Abdelnasser, M.~Youssef, and K.~A. Harras, ``{WiGest}: A ubiquitous
  {WiFi}-based gesture recognition system,'' in {\em Proc. of IEEE INFOCOM},
  pp.~1472--1480, April 2015.

\bibitem{Qian17Inferring}
K.~Qian, C.~Wu, Z.~Zhou, Y.~Zheng, Z.~Yang, and Y.~Liu, ``Inferring motion
  direction using commodity wi-fi for interactive exergames,'' in {\em Proc. of
  CHI Conference on Human Factors in Computing Systems}, pp.~1961--1972, ACM,
  2017.

\bibitem{Ali15Keystrode}
K.~Ali, A.~X. Liu, W.~Wang, and M.~Shahzad, ``Keystroke recognition using
  {WiFi} signals,'' in {\em Proc. of the 21st Annual International Conference
  on Mobile Computing \& Networking}, pp.~90--102, ACM, 2015.

\bibitem{Pu13Whole}
Q.~Pu, S.~Gupta, S.~Gollakota, and S.~Patel, ``Whole-home gesture recognition
  using wireless signals,'' in {\em Proc. of the 19th Annual International
  Conference on Mobile Computing \& Networking}, pp.~27--38, ACM, 2013.

\bibitem{Wang16WiHear}
G.~Wang, Y.~Zou, Z.~Zhou, K.~Wu, and L.~M. Ni, ``We can hear you with
  {Wi-Fi}!,'' {\em IEEE Transactions on Mobile Computing}, vol.~15,
  pp.~2907--2920, Nov 2016.

\bibitem{Wang14Eeyes}
Y.~Wang, J.~Liu, Y.~Chen, M.~Gruteser, J.~Yang, and H.~Liu, ``{E-eyes}:
  Device-free location-oriented activity identification using fine-grained wifi
  signatures,'' in {\em Proc. of the 20th Annual International Conference on
  Mobile Computing \& Networking}, pp.~617--628, ACM, 2014.

\bibitem{Wang15Understanding}
W.~Wang, A.~X. Liu, M.~Shahzad, K.~Ling, and S.~Lu, ``Understanding and
  modeling of {WiFi} signal based human activity recognition,'' in {\em Proc.
  of the 21st Annual International Conference on Mobile Computing \&
  Networking}, pp.~65--76, ACM, 2015.

\bibitem{Han14WiFall}
Y.~Wang, K.~Wu, and L.~M. Ni, ``{WiFall}: Device-free fall detection by
  wireless networks,'' {\em IEEE Transactions on Mobile Computing}, vol.~16,
  pp.~581--594, Feb 2017.

\bibitem{Sun15WiDraw}
L.~Sun, S.~Sen, D.~Koutsonikolas, and K.-H. Kim, ``{WiDraw}: Enabling
  hands-free drawing in the air on commodity {WiFi} devices,'' in {\em Proc. of
  the 21st Annual International Conference on Mobile Computing \& Networking},
  pp.~77--89, ACM, 2015.

\bibitem{Adib13See}
F.~Adib and D.~Katabi, ``See through walls with {WiFi}!,'' {\em SIGCOMM Comput.
  Commun. Rev.}, vol.~43, pp.~75--86, Aug. 2013.

\bibitem{Seifeldin13Nuzzer}
M.~Seifeldin, A.~Saeed, A.~E. Kosba, A.~El-Keyi, and M.~Youssef, ``Nuzzer: A
  large-scale device-free passive localization system for wireless
  environments,'' {\em IEEE Transactions on Mobile Computing}, vol.~12,
  pp.~1321--1334, July 2013.

\bibitem{Chen17TRBreath}
C.~Chen, Y.~Han, Y.~Chen, H.~Q. Lai, F.~Zhang, B.~Wang, and K.~J.~R. Liu,
  ``{TR-BREATH}: Time-reversal breathing rate estimation and detection,'' {\em
  IEEE Transactions on Biomedical Engineering}, vol.~PP, no.~99, pp.~1--14,
  2017.

\bibitem{Xu17TRIEDS}
Q.~Xu, Y.~Chen, B.~Wang, and K.~J.~R. Liu, ``{TRIEDS}: Wireless events
  detection through the wall,'' {\em IEEE Internet of Things Journal}, vol.~4,
  pp.~723--735, June 2017.

\bibitem{Adib14WiTrack}
F.~Adib, Z.~Kabelac, D.~Katabi, and R.~C. Miller, ``{3D} tracking via body
  radio reflections,'' in {\em 11th {USENIX} Symposium on Networked Systems
  Design and Implementation}, pp.~317--329, {USENIX} Association, 2014.

\bibitem{murphy2006design}
P.~Murphy, A.~Sabharwal, and B.~Aazhang, ``Design of warp: A flexible wireless
  open-access research platform,'' in {\em Proc. EUSIPCO}, pp.~53--54, 2006.

\bibitem{Hill09Electromagnetic}
D.~A. Hill, {\em Electromagnetic fields in cavities: deterministic and
  statistical theories}, vol.~35.
\newblock John Wiley \& Sons, 2009.

\bibitem{Tse05Fundamentals}
D.~Tse and P.~Viswanath, {\em Fundamentals of wireless communication}.
\newblock Cambridge university press, 2005.

\bibitem{Chen17Achieving}
C.~Chen, Y.~Chen, Y.~Han, H.~Q. Lai, F.~Zhang, and K.~J.~R. Liu, ``Achieving
  centimeter-accuracy indoor localization on {WiFi} platforms: A multi-antenna
  approach,'' {\em IEEE Internet of Things Journal}, vol.~4, pp.~122--134, Feb
  2017.

\bibitem{Zhang17time}
F.~Zhang, C.~Chen, B.~Wang, H.~Q. Lai, and K.~J.~R. Liu, ``A time-reversal
  spatial hardening effect for indoor speed estimation,'' in {\em Proc. of IEEE
  ICASSP}, pp.~5955--5959, March 2017.

\bibitem{Chiueh12Baseband}
T.-D. Chiueh, P.-Y. Tsai, and I.-W. Lai, {\em Baseband receiver design for
  wireless MIMO-OFDM communications}.
\newblock John Wiley \& Sons, 2012.

\bibitem{Shumway10Time}
R.~H. Shumway and D.~S. Stoffer, ``Time series analysis and its applications
  with r examples,'' {\em Time series analysis and its applications with R
  examples}, 2006.

\bibitem{Nee97Delay}
R.~Van~Nee, ``Delay spread requirements for wireless networks in the 2.4 {GHz}
  and 5 {GHzi} bands,'' {\em IEEE}, vol.~802, pp.~802--22, 1997.

\bibitem{Cleveland79}
W.~S. Cleveland, ``Robust locally weighted regression and smoothing
  scatterplots,'' {\em Journal of the American statistical association},
  vol.~74, no.~368, pp.~829--836, 1979.

\bibitem{Scheffe99ANOVA}
H.~Scheffe, {\em The analysis of variance}, vol.~72.
\newblock John Wiley \& Sons, 1999.

\bibitem{Kim09l1}
S.-J. Kim, K.~Koh, S.~Boyd, and D.~Gorinevsky, ``l1 trend filtering,'' {\em
  SIAM review}, vol.~51, no.~2, pp.~339--360, 2009.

\bibitem{Kozlov15Persistence}
Y.~Kozlov and T.~Weinkauf, ``{Persistence1D}: Extracting and filtering minima
  and maxima of 1d functions,'' {\em
  http://people.mpi-inf.mpg.de/weinkauf/notes/persistence1d.html}, pp.~11--01,
  2015.

\bibitem{Bagala12Evaluation}
F.~Bagal{\`a}, C.~Becker, A.~Cappello, L.~Chiari, K.~Aminian, J.~M. Hausdorff,
  W.~Zijlstra, and J.~Klenk, ``Evaluation of accelerometer-based fall detection
  algorithms on real-world falls,'' {\em PloS one}, vol.~7, no.~5, p.~e37062,
  2012.

\bibitem{Papoulis02Probability}
A.~Papoulis and U.~Pillai, {\em Probability, random variables, and stochastic
  processes}.
\newblock McGraw-Hill, 2002.

\end{thebibliography}

\end{document}